\documentclass[a4paper,11pt]{article}

\usepackage{jheppub} 

\usepackage[T1]{fontenc} 

\usepackage{bm}
\usepackage{latexsym}
\usepackage{dcolumn}
\usepackage{amsmath,amsfonts,amssymb, amsthm}
\usepackage{graphicx,epsfig}
\usepackage{environ}
\usepackage{mathtools}
\usepackage{braket}
\usepackage{float}
\usepackage{mathrsfs}

\usepackage{amsmath}
\usepackage{tensor}
\usepackage{fancyhdr}
\usepackage{subcaption}
\usepackage{hyperref}
\usepackage{graphicx,epstopdf}
\usepackage{capt-of}
\makeatletter
\newcommand*{\rom}[1]{\expandafter\@slowromancap\romannumeral #1@}
\makeatother

\hypersetup{
	colorlinks   = true, 
	urlcolor     = blue, 
	linkcolor    = blue, 
	citecolor   = red 
}

\def\b{\beta}
\def\g{\gamma}

\def\be{\begin{equation}}
\def\ee{\end{equation}}
\def\beq{\begin{eqnarray}}
\def\eeq{\end{eqnarray}}
\def\bs{\boldsymbol}

\newcommand{\levicivita}{}
\def\levicivita#1#{\tensor#1{\epsilon}}

\title{Gravity dual of Navier-Stokes equation in a rotating frame through parallel transport}
\author[a]{Sumit Dey,}
\author[b]{Shounak De}
\author[a]{and Bibhas Ranjan Majhi}

\affiliation[a]{Department of Physics, Indian Institute of Technology Guwahati, Guwahati 781039, Assam, India}
\affiliation[b]{Department of Applied Mathematics and Theoretical Physics, University of Cambridge, Wilberforce Road, Cambridge CB3 0WA, UK} 
 
\emailAdd{dey18@iitg.ac.in}
\emailAdd{sd868@cam.ac.uk}
\emailAdd{bibhas.majhi@iitg.ac.in}
\abstract{The fluid-gravity correspondence documents a precise mathematical map between a class of dynamical spacetime solutions of the Einstein field equations of gravity and the dynamics of its corresponding dual fluid flows governed by the Navier-Stokes (NS) equations of hydrodynamics. This striking connection has been explored in several dynamics-based approaches and has surfaced in various forms over the past four decades. In a recent construction, it has been shown that the manifold properties of geometric duals are in fact intimately connected to the dynamics of incompressible fluids, thus bypassing the conventional \textit{on-shell} standpoints. Following such a prescription, we construct the geometrical description that effectively captures the dynamics of an incompressible NS fluid with respect to a uniformly rotating frame. We propose the gravitational dual(s) described by bulk metric(s) in $(p+2)$-dimensions such that the equations of \textit{parallel transport} of an appropriately defined bulk velocity vector field when projected onto an induced timelike hypersurface require that the incompressible NS equation of a fluid relative to a uniformly rotating frame be satisfied at the relevant perturbative order in $(p+1)$-dimensions. We argue that free fluid flows on manifold(s) described by the proposed metric(s) can be effectively considered as an \textit{equivalent theory of non-relativistic viscous fluid dynamics with respect to (w.r.t) a uniform rotating frame}. We also present suggestive insights as to how space-time rotation parameters encode information pertaining to the inertial effects in the corresponding fluid dual.} 

\begin{document}
\maketitle
\flushbottom
\noindent

\section{\label{intro}INTRODUCTION}
The non relativistic incompressible Navier-Stokes (NS) equation \cite{Landau,Lautrup}
\begin{align}
\dot{\vec{v}} + \vec{v} . \nabla \vec{v} + \vec{\nabla} P - \eta \nabla^2 \vec{v} = 0 \,,
\label{0.1}
\end{align}
and the Einstein field equations of gravity
\begin{align}
G_{a b} = \frac{8 \pi G}{c^4}T_{a b} \,,
\label{0.2}
\end{align}
are two of the most important and well-studied differential equations in physics and mathematics. While the incompressible NS equation (\ref{0.1}) universally governs the dynamics of fluids in the hydrodynamic limit, the Einstein equation (\ref{0.2}) is known to universally govern the long-distance dynamics of gravitating systems. A precise mathematical bridge between (\ref{0.1}) and (\ref{0.2}) and their solution spaces is well documented by the fluid-gravity correspondence \cite{Hubeny:2010wp,Rangamani:2009xk}. This striking connection relating the dynamics of gravity to those of fluid equations, has gradually taken shape and surfaced in various forms over the past four decades. 

One of the earliest works relating the dynamics of gravity and that of hydrodynamics appeared in the doctoral thesis of Damour \cite{Damour:1979wya}, wherein there are suggestions of a connection between horizon and fluid dynamics. This work contains an expression now known as the Damour-Navier-Stokes (DNS) equation and is known to govern the geometric data on any null surface. The same equation is also obtained in terms of coordinates adapted to a null surface \cite{Padmanabhan:2010rp} \cite{Gourgoulhon:2005ng} by projecting the Einstein's equations of motion onto the null hypersurface (a similar analysis has also been done in \cite{Bhattacharya:2020wdl} for scalar-tensor gravity theory to obtain DNS like equation). Moreover, a corresponding action formulation of the same has been greatly detailed in \cite{Kolekar:2011gw}. A connection in this regard has also been obtained in the membrane paradigm approach by Price and Thorne in \cite{Price:1986yy}. The membrane paradigm was applied in \cite{Eling:2009pb} in the context of asymptotically AdS spacetimes to show the dynamics of the membrane being described by the incompressible NS equation. In \cite{Gourgoulhon:2008pu}, the authors have obtained an analogous DNS type equation for both future outer trapped horizons and dynamical horizons (which are spacelike). One peculiarity of the DNS equation as obtained on a null horizon is that the bulk viscosity of the horizon fluid is negative. This makes the null horizon fluid unfit to have a connection with ordinary fluids. However the authors of \cite{Gourgoulhon:2008pu} show that the horizon fluid on both future outer trapped horizons and the dynamical horizons have a positive value of the bulk viscosity. In the AdS/CFT context, it has been shown that the dissipative behaviour of an AdS black hole agrees with the hydrodynamics of the holographically dual CFT. In this approach the NS equation together with its corrections arise under a gradient expansion of the Einstein's equations. This has been studied extensively and important works in this regard include \cite{Bhattacharyya:2008kq, Bhattacharyya:2008jc, Bhattacharyya:2008ji, Policastro:2002se}. More recently in a cut-off surface approach by Bredberg et al. \cite{Bredberg:2011jq}, it has been shown by explicit construction that for every solution of the incompressible NS equation in $(p+1)$-dimensions, there is a uniquely associated dual solution of the vacuum Einstein equations in $(p+2)$-dimensions. The metric of \cite{Bredberg:2011jq}  has been extended to all orders perturbatively via gradient expansion in \cite{Compere:2011}, thus yielding higher order corrections to the NS equation as well as the incompressibility condition. In \cite{Brattan:2011my}, the authors have generalized the cut-off surface approach by expounding on the dynamics of the dual field theory living on the boundary of AdS spacetime, provided the Dirichlet boundary conditions on the $r=r_c$ cut-off surface is ensured. The authors show that there exists a critical radius as we go towards the horizon, beyond which, a relativistic description of the fluid living on the cut-off surface is not valid because of the acausal propagation of sound modes. Allowing the non relativistic scaling, the authors retrieve that Ricci flat gravitational duals to the incompressible NS equations. In \cite{Pinzani-Fokeeva:2014cka}, the authors provide a general approach to fluid/gravity correspondence, where the base metric is no longer the flat Rindler metric, but rather a generic static metric. The spacetime is endowed with a general bulk stress energy tensor and an event horizon. This cut-off surface approach has been applied in various cases, see \cite{Huang:2011he, De:2018zxo, Zhang:2012uy}. For example, it was extended for higher curvature gravity theories \cite{Chirco:2011ex, Bai:2012ci, Cai:2012mg, Zou:2013ix, Hu:2013lua} as well as for the AdS \cite{Cai:2011xv, Huang:2011kj} and dS \cite{Anninos:2011zn} gravity theories (for other theories, like black branes, see \cite{Ling:2013kua}). Very recently, two of the authors of this paper showed in \cite{De:2018zxo} that an incompressible DNS-like equation can be obtained in the cut-off surface approach. In this case the obtained metric is a solution of Einstein's equations of motion in the presence of a particular type of matter. Also a corresponding relativistic situation has been discussed extensively in \cite{Eling:2012ni}. Symmetries of the vacuum Einstein equations have been exploited to develop a formalism for solution generating transformations of the corresponding NS fluid duals in \cite{Berkeley:2012kz}. The fluid description on the Kerr horizon has also been explored extensively in \cite{Lysov:2017cmc} (see \cite{Wu:2013aov} for the isolated horizon case). The correspondence has also been established for general rotating black holes yielding a Coriolis force term \cite{Wu:2015pzg}. For extensive reviews of the fluid-gravity correspondence, refer to \cite{Padmanabhan:2009vy, Hubeny:2010wp, Rangamani:2009xk}.

Having detailed out the conventional approaches to this fascinating connection relating the dynamics of gravity to that of hydrodynamics, a novel interpretation of the same correspondence has been established in a new setting. In a recent work by the authors of this paper \cite{De:2019wok}, a new formalism was established to understand the fluid-gravity correspondence from a different standpoint. In the previous cut-off surface approach the underlying physics is that there exists a non trivial map between the fluid side and the gravity side constrained by their dynamical equations of motion. This approach lays out the connection or duality between the dynamics of the incompressible fluid and that of the Einstein's gravitational equations of motion via the conservation of the Brown-York stress tensor on the gravity side. However fundamentally, the physics in \cite{De:2019wok} is quite different. Here the correspondence is between an incompressible fluid living in Minkowski spacetime and that of appropriately defined bulk velocity field in curved spacetime. We then encode the dynamics of the bulk velocity field congruence in order to have a map between the fluid in the Minkowski spacetime and the bulk velocity field in the curved manifold. By dynamics, we impose that the acceleration of the congruence of  bulk velocity field on the $r=r_c$ timelike hypersurface is zero i.e the bulk velocity congruence is parallel transported on the $r=r_c$ slice. This allows us to have a map between the dynamics on both sides. The incompressible NS equation of the fluid is mapped to a "free" bulk velocity congruence on the $r=r_c$ hypersurface. The essence of the physics in \cite{De:2019wok} is that dynamics of the incompressible viscous fluid in Minkowski spacetime can be studied as the dynamics of a "free" parallel transported bulk velocity field on the cut-off slice. As a result of the projection of the parallel transport as being the analogue of the dynamics on the manifold side, all the dynamical degrees of freedom of the fluid are encoded in the manifold properties of the spacetime. The constraint of the incompressibility condition on the fluid side is shown to naturally arise from the vanishing of the expansion parameter corresponding to the bulk velocity field. It is for this reason why the projection of the parallel transport equation of the bulk velocity field on the cut-off slice is so important in this framework. Moreover this mapping between the two sides bypasses the Einstein's field equations as an added advantage and hence is an \textit{off-shell} duality between the incompressible fluid dynamics and parallel transport dynamics of a bulk velocity on the cut-off hypersurface. This approach to the fluid-gravity correspondence is completely different from existing works in this direction. This work \cite{De:2019wok} essentially forms the basis of the current paper which attempts to construct a gravitational dual of the incompressible NS equation in a rotating frame, the details of which are discussed in the following paragraphs.

The \textit{notations} used throughout the paper are clarified as follows. All lowercase Latin letters denote the bulk spacetime coordinate indices and run from $a, b = 0, \dots, p+1$. The uppercase Latin letters denote the transverse coordinates intrinsic to the hypersurface (i.e., the angular sector of the metric) and they run from $A, B = 2, \dots, p+1$ as the labels $0$ and $1$ have been chosen for time and radial coordinates, respectively. 

\section{OUTLINE OF THE PAPER : A brief review on incompressible NS w.r.t a rotating frame and motivation}\label{outline}
The incompressible NS equation in an inertial frame of reference (\ref{0.1}) takes the following form in index notation
\begin{align}
\partial^{'}_{\tau} v^{'}_A + v^{'C} \partial^{'}_C v^{'}_A + \partial^{'}_{A} P -\eta \partial^{'2} v^{'}_A &= 0 \nonumber \\
\partial^{'}_{A} v^{'A} &= 0
\label{0.3}
\end{align}
where the $v^{'}_A$ denotes the fluid velocity with respect to an inertial coordinate system $(\tau^{'}, \{x^{'A}\})$. All the derivatives are with respect to this primed inertial coordinate system. However for a fluid relative to the rotating frame, inertial effects (like the Coriolis and the centrifugal forces) need to be accounted for in the above equation when dealing with the dynamics of the fluid.
The coordinates of the fluid element with respect to the rotating frame are designated as $(\tau, \{x^A\}).$ The transformations of the position vectors, the velocities and the accelerations of the fluid element between the rotating frame and the inertial frame are given respectively via the relations,
\begin{align}
\boldsymbol{x}  &= R \cdot \boldsymbol{x^{'}} \nonumber \\
\boldsymbol{v} &= \boldsymbol{R}\cdot\boldsymbol{v^{'}} - \boldsymbol{\Omega}\times \boldsymbol{x} \nonumber \\
\boldsymbol{a} &= \boldsymbol{R} \cdot \boldsymbol{a^{'}} - \boldsymbol{\Omega} \times (\boldsymbol{\Omega} \times \boldsymbol{x}) - 2(\boldsymbol{\Omega}\times \boldsymbol{v}) ~,
\end{align}
where $\boldsymbol{\Omega}$ is the uniform angular velocity of the rotating frame. $\boldsymbol{R}$ denotes the general time dependent rotation matrix about any arbitrary plane. The centrifugal acceleration is given via $\boldsymbol{\Omega} \times (\boldsymbol{\Omega} \times \boldsymbol{x})$ and the Coriolis acceleration via $2 (\boldsymbol{\Omega}\times \boldsymbol{v})$. Thus the incompressible NS equation relative to a uniformly rotating frame for a  non relativistic viscous fluid system (with no external forces) can be written as,
\begin{align}
\partial_{\tau} v_{A} + v^{B}\partial_{B} v_{A} + \partial_{A}P - \eta \partial^2 v_{A} &= -2 (\boldsymbol{\Omega}\times \boldsymbol{v})_{A} - \Big(\boldsymbol{\Omega} \times (\boldsymbol{\Omega} \times \boldsymbol{x})\Big)_{A}~; 
\nonumber \\
\partial_{A}v^{A} &= 0 ~,
\label{rotationNS1}
\end{align}
where the Coriolis force is identified as $-2 (\boldsymbol{\Omega}\times \boldsymbol{v})$ and the centrifugal force as $-\boldsymbol{\Omega} \times (\boldsymbol{\Omega} \times \boldsymbol{x})$.

The centrifugal force term can be identified as the gradient of a certain centrifugal potential (on the assumption that the origin of the rotating coordinate system lies on the axis of rotation) which can then be incorporated inside the dynamical fluid pressure $P$ to identify an effective pressure term $P_{\textrm{eff}}$  \cite{Landau}. The effective pressure $P_\textrm{eff}$ can then be identified as,
\begin{eqnarray}
P_{\textrm{eff}} = P - \frac{1}{2}|\boldsymbol{\Omega} \times \boldsymbol{x}|^2 ~.
\label{0.4}
\end{eqnarray}
Thus, another form of the NS equation in the rotating frame is 
\begin{align}
\partial_{\tau} {v}_A + {v}^C \partial_C {v}^A + \partial_A P_{\text{eff}} -\eta \partial^2 {v}_A &= -2 (\boldsymbol{{\Omega}} \times \boldsymbol{v})_A \nonumber \\
\partial_{A}v^{A} &= 0 ~.
\label{0.5}
\end{align}
The details of this derivation can be followed from \cite{Landau} and \cite{Lautrup}.

In the above we found that the NS equation can be casted in two forms (\ref{rotationNS1}) and (\ref{0.5}). Here, we attempt to construct the gravitational dual of both of them, in the manifold based approach to the fluid-gravity correspondence as established in \cite{De:2019wok}. Our main target will be to construct  metrics which will lead to the NS equations when the parallel transport equation of a suitably chosen velocity vector is projected on the timelike hypersurface of  these metrics.  We first concentrate on finding dual of Eq. (\ref{rotationNS1}) and then for Eq. (\ref{0.5}). We shall observe that, although the above two equations represent the same NS equation, the manifolds are distinctly different. However the formalism of \cite{De:2019wok} on both these distinct metrics yield equivalently the NS equations (w.r.t the rotating frame) for the fluid dual.

Our basic organization of the paper is roughly as follows. In section \ref{sectionA} and \ref{sssection1} respectively, we begin by proposing two different bulk metrics in $(p+2)$-dimensions on which we consider the equations of the projection of the parallel transport of an appropriately defined velocity vector field. We then show in subsections \ref{ssection2} and \ref{sssection2} that the projection of the parallel transport equations onto a timelike induced hypersurface require that the fluid dynamical NS equations(w.r.t the rotating frame) be satisfied in $(p+1)$-dimensions. In subsections \ref{ssection3} and \ref{sssection3} we then show that the incompressibility condition of the fluid as viewed from the rotating frame derives from a vanishing expansion parameter $\theta$ when projected onto the same timelike induced hypersurface. We also identify the connections between the rotation parameter(s) on the gravity side and those in its fluid counterpart. The next section \ref{sectionB} shares further insights into the construction of the two proposed bulk metrics which are genuinely curved backgrounds. By doing so, we show in this process. that the present considered proposed metrics are in no way related by a diffeomorphism of the metric presented in \cite{De:2019wok}. In the final section \ref{sectionC}, we discuss the consequences of the two different gravity/metric duals we obtain for a NS fluid (relative to a rotating frame) in this parallel transport framework of the fluid-gravity correspondence and offer a few concluding remarks. The computation of the inverse(s) of the $(p+2)$-dimensional bulk metric(s), the relevant connection coefficients, the order-wise calculations of the projected parallel transport equations and that of the expansion parameter $\theta$ are explicitly detailed out in the appendices \ref{appA}, \ref{appA1}, \ref{appB}, \ref{appB1}, \ref{appC} and \ref{appC1}.


\section{SCALE INVARIANCE OF THE FLUID DYNAMICAL EQUATIONS}\label{scaleinvariance}
We  will propose two different metrics that effectively capture the dynamics of the viscous, NS fluid system relative to a uniformly rotating frame.  Without using the Einstein field equations, we will simply project the acceleration of an appropriately chosen bulk velocity congruence onto a specific chosen hypersurface for both of these metrics and demand it to be zero. Our main results, the specific forms of the manifolds, will be obtained based on the scaling invariance of the NS equation. Therefore, in the below, we start the section by a discussion on this.

The incompressible NS equation $\eqref{0.1}$ satisfies a well known scaling symmetry that is briefly stated as follows. If the solution space $(v^A,P)$ of the incompressible NS equation is scaled down by a certain hydrodynamic parameter $\epsilon$ as,
\begin{eqnarray}
v_{A}^{\epsilon}(x^A,\tau) = \epsilon v_A(\epsilon x^A,\epsilon^2 \tau); \quad P^{\epsilon}(x^A,\tau) = \epsilon^2 P(\epsilon x^A,\epsilon^2 \tau)
\label{scaling1}
\end{eqnarray}
then the NS equation remains invariant under the above scaling transformations. A detailed derivation of the scale invariance of the NS equation can be found in Appendix A of \cite{De:2018zxo}. The incompressible NS equation for a viscous fluid w.r.t the rotating frame ($\eqref{rotationNS1}$ or equivalently $\eqref{0.5}$) also remains scale invariant if we identify that the uniform angular velocity $\boldsymbol{\Omega}$ scales as 
\begin{eqnarray}
\Omega_{A}^{\epsilon} = \epsilon^2 \Omega_{A} ~.
\label{scalingonOmega}
\end{eqnarray}
The justification of this comes from the fact that angular velocity has the dimensions of the inverse of time $\tau$. Since for the scale invariance, $\tau$ scales as order $\mathcal{O}(\epsilon^{-2})$, hence the components of the angular velocity scale as order $\mathcal{O}(\epsilon^2)$. 
Thus via the hydrodynamic scaling $\epsilon$ we can generate a class of solutions parametrized by $(v^{\epsilon}_{A}, P^{\epsilon})$. The hydrodynamic scalings of the dynamical variables, the constant angular velocity components along the spatial and temporal derivatives follow as,
\begin{align}
v_{A} \sim \mathcal{O}(\epsilon), \quad P \sim \mathcal{O}(\epsilon^2),\quad \Omega_{A} \sim \mathcal{O}(\epsilon^2),\nonumber \\
\partial_A \sim \mathcal{O}(\epsilon), \quad \partial_{\tau} \sim \mathcal{O}(\epsilon^2) ~.
\label{scalingssummary}
\end{align}

\section{FLUID DYNAMICS VIA PARALLEL TRANSPORT: Choice \rom{1}}\label{sectionA}
For the proposed metric we follow the methodology applied in \cite{Bredberg:2011jq} and lay out the metric order by order in terms of the hydrodynamic scaling parameter $\epsilon$.

We propose a metric of the form
\begin{eqnarray}
ds^2_{p+2}  = g_{ab}dx^a dx^b && =  -r d\tau^2 + 2 d\tau dr + dx_Adx^A \nonumber \\ 
&&-2\b_A f(r) d\tau dx^A -2 \b_A g(r) dr dx^A \nonumber \\
&& + \Big(a_3 (x^{A}\Omega_{A}x^{B}\Omega_{B}) + a_4 (\delta^{AB}\Omega_{A}\Omega_{B}\delta_{CD}x^{C}x^{D})\Big)d\tau^2 \nonumber \\
&&- \Big(\frac{2a_1}{r_c}\partial_A P +2 a_2 \partial^2v_A -\frac{4}{r_c}f(r)v^D \partial_D \b_A\Big)dx^A dr \nonumber \\
&& + \mathcal{O}(\epsilon^4)  ~.
\label{metrixproposal}
\end{eqnarray}
The metric is written in such  a way that the first line is of the order $\mathcal{O}(\epsilon^0)$ and is simply the flat Rindler metric written in ingoing Eddington-Finkelstein coordinates, the second line is the order $\mathcal{O}(\epsilon^1)$,the third line is of order $\mathcal{O}(\epsilon^2)$ and the fourth line is of the order $\mathcal{O}(\epsilon^3)$. $a_1$, $a_2$, $a_3$ and $a_4$ serve as constants whose values will be fixed later. We impose as a condition that $\b_A$ scales as order $\epsilon^1$. The present metric matches with the proposed metric in \cite{De:2019wok} (with $a_3$ set to zero) if we set the constants here $a_3$ and $a_4$ as well as $\b_A$ equal to zero.
The metric expanded to order $\mathcal{O}(\epsilon^2)$ serves as the base metric upon which the perturbation at order $\mathcal{O}(\epsilon^3)$ has been added. Initially, at this metric level, the set $\{\Omega_A\}$ are just some uniform components that we demand to scale as order $\mathcal{O}(\epsilon^2)$. No identification of $\{\Omega_A\}$   at this point can be made with the overall uniform angular velocity components for the fluid to be described. Same goes for the set $\{\b_A\}$. Similarly, at the metric level, the fields $P(\tau, x^A)$ and $\{v_A(\tau,x^B)\}$ are not to be initially identified with the pressure perturbations and the velocity components of the fluid. All that we require at this metric level is that the fields $P(\tau, x^A)$ and $\{v_A(\tau,x^B)\}$ scale as order $\mathcal{O}(\epsilon^2)$ and $\mathcal{O}(\epsilon^1)$ respectively in terms of the hydrodynamic scaling parameter $\epsilon$. The analogy/correspondence will emerge only after the formal machinery of the projection of the parallel transport equation has been applied.

The physical interpretation behind the construction of such a metric is explained now. We consider the effect of the following coordinate transformation on a  $4D$ Minkowski metric $ds^2 = \eta_{ab}dx^a dx^b$ with $\eta$ $\coloneqq $ diag$\{-1,+1,+1,+1\}$ and $\{x^a\}$ $\coloneqq$ $\{t,x,y,z\}$,
\begin{align}
\tilde{t} &= t \nonumber \\
\tilde{x} &= x \cos(\omega t) + y \sin(\omega t) \nonumber \\
\tilde{y} &= -x\sin(\omega t) + y \cos(\omega t) \nonumber \\
\tilde{z} &= z ~.
\end{align}
Under the effect of the following coordinate transformation where $\omega$ is a constant, the Minkowski metric becomes,
\begin{eqnarray}
ds^2 = -\Big[1- \omega^2(\tilde{x}^2+\tilde{y}^2)\Big]d\tilde{t}^2 + 2 \omega(\tilde{x}d\tilde{y}-\tilde{y}d\tilde{x})d\tilde{t} + d\tilde{x}^2 + d\tilde{y}^2 + d\tilde{z}^2 ~.
\end{eqnarray}
It can be verified that in the non relativistic limit, the geodesic equation of motion in this manifold reduces to the usual Newton's laws for a free particle in a rotating frame, incorporating both the Coriolis and the centrifugal effects. Taking hints from the above construction we see that there ought to be coefficients of $d \tau d x^A$ in our proposed metric that are of the order $\mathcal{O}(\epsilon)$ and so should be terms of order $\mathcal{O}(\epsilon^2)$ containing $d\tau^2$. In our proposed metric, $\beta_A$ is, at this level a function of $(\tau,x^A )$. However $\Omega$ in our proposed metric is uniform and hence independent of $\tau$, $r$ and $\{x^A\}$. It must be reminded that this is just a formal analogical way of proposing the present metric \eqref{metrixproposal}. 
A pertinent question might appear as to the emergence of the extra parameter $\beta_A$ in the proposed metric \eqref{metrixproposal}. It will be shown later after the formal machinery (of parallel transport) has been applied, that $\{\Omega_A\}$ and $\{\b_A\}$ will be demanded to be dependent on each other for consistency.  

We assume for the moment that in \eqref{metrixproposal} $f(r)$ and $g
(r)$ are smooth functions of only the radial coordinate with the imposition that $f(r= r_c) \neq 0$ and $g(r=r_c) \neq 0$. Here we also mention that $r = r_c$ is the location of the timelike hypersurface (in the bulk manifold) that we are interested in. The location is at any finite distance between the horizon $r = 0$ and radial infinity. 
The \textit{base bulk metric} here is 
\begin{eqnarray}
g^{(0)}_{ab}dx^{a}dx^{b} = -r d\tau^2 + 2 d\tau dr + dx_Adx^A -2\b_A f(r) d\tau dx^A -2\b_A g(r)drdx^A \nonumber \\
+ \Big(a_3 (x^{A}\Omega_{A}x^{B}\Omega_{B}) + a_4 (\delta^{AB}\Omega_{A}\Omega_{B}\delta_{CD}x^{C}x^{D})\Big)d\tau^2
\label{basemetrix}
\end{eqnarray}
over which the \textit{perturbation}
\begin{eqnarray}
h^{(3)}_{ab} dx^{a}dx^{b} = - \Big(\frac{2a_1}{r_c}\partial_A P +2 a_2 \partial^2v_A -\frac{4}{r_c}f(r)v^D \partial_D \b_A\Big)dx^A dr
\label{perturbation3}
\end{eqnarray}
to the order $\epsilon^3$ has been applied. 
 We denote the perturbation as $h^{(3)}_{ab}$. As a result, the base bulk metric is curved, which can be checked by calculating the components of the Riemann curvature tensor $R^{a}_{~bcd}$. The fields $P(\tau, x^A)$ and $v^A(\tau, x^A)$  are independent of the radial coordinate $r$. We shall show that the perturbation $h^{(3)}_{ab}$ in the proposed metric  contains information about all the forcing terms in the Navier-Stokes equation (relative to the rotating frame) for a viscous incompressible fluid. This proposed metric acts as the metric/gravity dual to the nonrelativistic fluid dynamical equations written in a rotating coordinate system. Projecting the acceleration of the bulk fluid congruence in this given spacetime onto the timelike hypersurface $r = r_c$ and demaniding it to be zero, we obtain the corresponding fluid dynamical equation (in the process losing general covariance).
 

\subsection{Parallel transport of the velocity field} \label{ssection2}
All the kinematical and the dynamical quantities of interest to us (which are defined for the entire spacetime manifold) will be projected onto the $r= r_c$ timelike hypersurface.
The projection tensor onto the hypersurface $r = r_c$ is given by,
\begin{eqnarray}
\g_{ab} = g_{ab} - n_a n_b ~,
\end{eqnarray}
where $n_a$ is the unit normal on this hypersurface. The hypersurface being timelike, its unit normal satisfies $n^an_a = +1$(spacelike unit normal). Since the base bulk metric is of the order $\mathcal{O}(\epsilon^2)$, the computation of $n_a$ (on the $r=r_c$ hypersurface) yields,
\begin{align}
n_{\tau}\Big|_{r=r_c} &= 0 \nonumber \\
n_r\Big|_{r=r_c} &= \frac{1}{\sqrt{r_c}} +\frac{1}{r_c^{3/2}}\Big\{a_3(x^A\Omega_A x^{B}\Omega_B) + a_4(\delta^{AB}\Omega_A\Omega_B \delta_{CD}x^Cx^D)\Big\} \nonumber \\
&-\delta^{AB}\beta_A \b_B(f^2 + 2rfg + r^2g^2) + \mathcal{O}(\epsilon^4) \nonumber \\
\{n_A\}\Big|_{r=r_c} &= 0 ~.
\end{align}
The calculation for the components of the normal to the hypersurface has been carried upto $\mathcal{O}(\epsilon^2)$. The components of the projection tensor (to order $\mathcal{O}(\epsilon^2)$) on the $r = r_c$ slice follow as 
\begin{eqnarray}
&& \g_{\tau \tau} = -r_c + a_3(x^A\Omega_A x^{B}\Omega_B) + a_4(\delta^{AB}\Omega_A\Omega_B \delta_{CD}x^Cx^D), \nonumber \\
&&\g_{\tau r} = 1, \quad \quad \g_{\tau A} = -\b_A f(r_c),\nonumber \\
&&\g_{rr} = -\frac{1}{r_c} - \frac{1}{r_{c}^{2}}\Big\{a_3(x^A\Omega_A x^{B}\Omega_B) + a_4(\delta^{AB}\Omega_A\Omega_B \delta_{CD}x^Cx^D) - \delta^{AB}\b_A \b_B(f^2 + 2rfg + r^2 g^2)\Big\},\nonumber \\
&&\g_{rA} = -\b_A g(r_c) ,\quad \quad \g_{AB} = \delta_{AB} ~.
\label{projectors}
\end{eqnarray}
Raising these covariant projectors via the inverse metric tensor $g^{ab}$ (see Appendix \ref{appA}), we obtain the contravariant components of the projection tensor to the hypersurface $r= r_c$ upto the order $\mathcal{O}(\epsilon^2)$ as,
\begin{eqnarray}
&& \g^{\tau \tau} = -\frac{1}{r_c} - \frac{1}{r_c^{2}}\Big\{a_3(x^A\Omega_A x^{B}\Omega_B) + a_4(\delta^{AB}\Omega_A\Omega_B \delta_{CD}x^Cx^D)\Big\} + \frac{f^2(r_c)}{r_c^{2}}\delta^{AB}\b_A \b_B, \nonumber \\
&& \g^{\tau r} = 0, \quad \quad \g^{\tau A} = -\frac{f(r_c)}{r_c}\delta^{AB}\b_{B}, \nonumber \\
&& \g^{rr} = 0, \quad \quad \g^{rA} = 0, \quad \quad \g^{AB} = \delta^{AB} +\frac{f^2(r_c)}{r_c} \delta^{AC}\delta^{BD}\b_C\b_D  ~.
\label{contraprojectors}
\end{eqnarray}
Now in the given spacetime manifold we define a bulk velocity field as $v^a = (1,0,v^A)$, such that there is no flow in the radial direction. The acceleration of the congruence of the velocity field is given via $a^i = v^b\nabla_b v^i$. We demand that the acceleration of this congruence as projected on the $r=r_c$ timelike hypersurface is zero. The component of the acceleration of the manifold fluid congruence on the $r=r_c$ slice is zero. This is represented by,
\begin{eqnarray}
\g_{ac} v^b \nabla_b v^a {|_{r=r_c}} = 0 ~.
\label{projectedautoparallel}
\end{eqnarray}
It is in this sense that we are calling \ref{projectedautoparallel} as a parallel transport equation or more correctly the projected parallel transport equation, i.e the fluid congruence is "free" only on the $r=r_c$ hypersurface.
Setting the free index $c$ as $\tau$, we obtain the left hand side (LHS) of  $\eqref{projectedautoparallel}$ on the hypersurface $r =r_c$ of \eqref{metrixproposal} as,
\begin{eqnarray}
\gamma_{a \tau}(v^b \nabla_b v^a)|_{r=r_c} &=& \g_{\tau \tau}\Big(\Gamma^{\tau}_{\tau \tau}+ 2\Gamma^{\tau}_{\tau A}v^A + \Gamma^{\tau}_{A B}v^A v^B\Big) + \g_{r \tau}\Big(\Gamma^{r}_{\tau \tau} + 2 \Gamma^{r}_{\tau A} v^A + \Gamma^{r}_{AB}v^A v^B\Big)\nonumber \\ 
&& \g_{B \tau}\Big(\partial_{\tau}v^B + v^A \partial_A v^B + \Gamma^{B}_{\tau \tau} + 2 \Gamma^{B}_{\tau A} v^A + \Gamma^{B}_{A D}v^A v^D\Big)|_{r=r_c} ~.
\label{projautoattau}
\end{eqnarray}
The evaluation of this equation $\eqref{projautoattau}$ yields zero at orders $\mathcal{O}(\epsilon^0)$, $\mathcal{O}(\epsilon^1)$, $\mathcal{O}(\epsilon^2)$ and $\mathcal{O}(\epsilon^3)$, which has been detailed out in Appendix \ref{appB}. Hence the projected parallel transport equation on the $r=r_c$ hypersurface with the free index $c = \tau$  is trivially satisfied upto order $\mathcal{O}(\epsilon^3)$.

We next turn our attention to the free index $c = r$ and have the following L.H.S of \eqref{projectedautoparallel} as,

\begin{eqnarray}
\gamma_{a r}(v^b \nabla_b v^a)|_{r=r_c} &=& \gamma_{\tau r}\Big(\Gamma^{\tau}_{\tau \tau} + 2\Gamma^{\tau
}_{\tau A}v^A + \Gamma^{\tau}_{A B}v^A v^B\Big) + \gamma_{rr}\Big(\Gamma^{r}_{\tau \tau} + 2\Gamma^{r}_{\tau A}v^A + \Gamma^{r}_{A B}v^A v^B\Big) \nonumber \\
&& \gamma_{A r}\Big(\partial_{\tau} v^A + v^B \partial_B v^A + \Gamma^{A}_{\tau \tau} + 2 \Gamma^{A}_{\tau B}v^B + \Gamma^{A}_{B C}v^B v^C\Big)|_{r =r_c} ~. 
\label{projauoatr}
\end{eqnarray}
Evaluating $\eqref{projauoatr}$ order by order, we see that it vanishes at $\mathcal{O}(\epsilon^0)$, $\mathcal{O}(\epsilon^1)$ and $\mathcal{O}(\epsilon^3)$. However at $\mathcal{O}(\epsilon^2)$,  $\eqref{projauoatr}$ yields a quantity proportional to  $\frac{f^2(r_c)}{r_c}+ 2 f(r_c)g(r_c)+r_{c} g^2(r_c)$. The details are listed in Appendix \ref{appB}. The imposition of $\eqref{projectedautoparallel}$ implies,
\begin{equation}
f^2(r_c) + 2 r_c f(r_c) g(r_c) + r^{2}_{c} g^{2}(r_c) = 0 ~,
\label{condition_on_g}
\end{equation}
the consequence of which we obtain $g(r_c) = -\frac{f(r_c)}{r_c}$, which has to be satisfied on the $r = r_c$ timelike slice.

At this point we find out the covariant components of the velocity field. The contravariant components were defined as $v^a = (1,0,{v}^A)$. The covariant components of the velocity field (lowered via the base bulk metric) are,
\begin{eqnarray}
&& v_{\tau} = -r -f \b_A v^A +a_3(x^A\Omega_A x^{B}\Omega_B) + a_4(\delta^{AB}\Omega_A\Omega_B \delta_{CD}x^Cx^D) + \mathcal{O}(\epsilon^4) ~, \nonumber \\
&& v_r = 1-g\b_A v^A + \mathcal{O}(\epsilon^4) ~, \quad {v}_A = -f\b_A + \delta_{AB}v^B  ~.
\end{eqnarray}

We finally look at the projection of the parallel transport equation \eqref{projectedautoparallel} with the free index $c = A$. As a consequence we obtain for the LHS of \eqref{projectedautoparallel},
\begin{eqnarray}
\gamma_{a A}(v^b \nabla_b v^a)|_{r=r_c} &=&\gamma_{\tau A}\Big(\Gamma^{\tau}_{\tau \tau} + 2 \Gamma^{\tau}_{\tau D} v^D + \Gamma^{\tau}_{CD}v^C v^D \Big) + \gamma_{rA}\Big(\Gamma^{r}_{\tau \tau} + 2 \Gamma^{r}_{\tau D}v^D + \Gamma^{r}_{CD}v^C v^D\Big) \nonumber \\
&& \gamma_{BA}\Big(\partial_{\tau}v^B + v^{C}\partial_{C}v^B + \Gamma^{B}_{\tau \tau}+ 2 \Gamma^{B}_{\tau D}v^D +\Gamma^{B}_{C D}v^C v^D\Big)~.
\label{projautoatA}
\end{eqnarray}
Evaluating \eqref{projautoatA} order by order, we see that it vanishes at $\mathcal{O}(\epsilon^0)$, $\mathcal{O}(\epsilon^1)$ and $\mathcal{O}(\epsilon^2)$. Evaluation of the above equation on the $r=r_c$ cut-off hypersurface at $\mathcal{O}(\epsilon^3)$, yields (see Appendix \ref{appB} for details),
\begin{eqnarray}
\mathcal{O}(\epsilon^3): \gamma_{a A}(v^b \nabla_b v^a)|_{r=r_c} &=& \partial_{\tau} v_A + v^C\partial_C v_A + f(r_c)v^C\partial_C\b_A  \nonumber \\
&& + \frac{r_c}{2}\Big(\frac{a_1}{r_c}\partial_A P + a_2 \partial^2 v_A - \frac{2}{r_c}f(r_c)v^D\partial_D\beta_A\Big)\nonumber \\
&&  + f(r_c)\Big(\partial_A \b_D -\partial_D \b_A \Big)v^D -a_3\Omega_A(\Omega_P x^P)\nonumber \\
&&- a_4(\delta^{CD}\Omega_C \Omega_D \delta_{AP}x^{P})~.
\label{projautoatA3}
\end{eqnarray}
The validity of \eqref{projectedautoparallel} with the free index $c = A$ at order $\mathcal{O}(\epsilon^3)$ imposes the requirement,
\begin{eqnarray}
\partial_{\tau} {v}_A+ {v}^C \partial_C {v}_A + \Big(\frac{a_1}{2}\partial_A P + \frac{a_2}{2} r_c \partial^2 {v}_A \Big) \nonumber \\
+ f(r_c)\Big(\partial_A \b_D - \partial_D \b_A \Big){v}^D-a_3\Omega_A(\Omega_P x^P) - a_4(\delta^{CD}\Omega_C \Omega_D \delta_{AP}x^{P}) &=& 0 ~. 
\label{asNavierStokes}
\end{eqnarray}
So finally having arrived at \eqref{asNavierStokes}, we want its correspondence to be made with \eqref{rotationNS1} ( where $\boldsymbol{{v}}$ denotes the velocity of the fluid element with respect to the rotating frame). It is at this level that we start making the necessary identifications between the quantities in the  $(p+2)$ dimensional manifold sector and the respective quantities in the $(p+1)$ dimensional fluid dynamics sector. In order to make this parallel evident, the following identifications need to made. The fields $P(\tau, x^A)$ and $\{v_A(\tau,x^B)\}$ in the manifold sector (that scaled as order $\mathcal{O}(\epsilon^2)$ and $\mathcal{O}(\epsilon)$ respectively) are indeed the pressure perturbation and the velocity field on the fluid sector. In the same way, we identify that the component $\Omega_A$ on the gravity side is equivalent to the uniform angular velocity of the frame that studies the fluid. We also identify that the uniform angular velocity components are related to the components $\{\b_A\}$ via,
\begin{eqnarray}
(\partial_A \b_B - \partial_B \b_A){v}^B = -(\bs{\Omega} \times \bs{{v}})_A ~,
\label{pde2}
\end{eqnarray}  
as a result of which we can write
\begin{eqnarray}
\boldsymbol{\Omega} = \bs{\nabla} \times \bs{\b} ~.
\label{Beta}
\end{eqnarray}
 We hence see that at the metric level the components $\{\beta_A\}$ are not free parameters, but rather constrained on $\{\Omega_A\}$.
We can further constrain the functional form of $\beta$ via the fact that \eqref{rotationNS1} represents the NS equation described in a uniformly rotating reference frame. Hence we demand that the functional form of $\boldsymbol{\beta}$ ought to be such that $\boldsymbol{\Omega}$ is a constant vector. One particular solution of \eqref{Beta} is 
\begin{eqnarray}
\beta_{D} = \frac{1}{(p-1)} \epsilon^{A}_{~~BD} \Omega_{A}x^{B} + \phi_D{(\tau)} ~,
\end{eqnarray}
where $\phi(\tau)$ is any arbitrary function of $\tau$. 
We then constrain the functional form of $f(r)$ on the cutoff hypersurface as $f(r_c) = -2$ as a result of which $g(r_c) = \frac{2}{r_c}$.
Identifying the constants $a_i$'s in the proposed metric as \eqref{metrixproposal} as,
\begin{eqnarray}
a_1 = 2, \quad \quad a_2 = -2, \quad \quad a_3 = -1,  \quad \quad a_4=  1 
\label{constants1} 
\end{eqnarray}
and the kinematic viscosity term $\eta$ as $\eta = r_c$, \eqref{asNavierStokes} formalizes to,
\begin{eqnarray}
\partial_{\tau} {v}_A + {v}^C \partial_C {v}_A + \partial_A P - \eta \partial^2 {v_A}+ 2(\levicivita{_{ABC}}\Omega^B {v}^C)\nonumber\\
 + \Omega_A(\Omega_P x^P) - (\delta^{CD}\Omega_C \Omega_D \delta_{AP}x^P)= 0 ~.
\label{NavierStokes}
\end{eqnarray}
It can be easily shown that the last two terms of the L.H.S of \eqref{NavierStokes} are exactly the centrifugal force component $\Big(\boldsymbol{\Omega} \times (\boldsymbol{\Omega}\times \boldsymbol{x})\Big)_A$.
So finally our formalism of the projection of the parallel transport equation on the timelike slice yields
\begin{equation}
\partial_{\tau} {v}_A + {v}^C \partial_C {v}_A + \partial_A P - \eta \partial^2 {v_A}+2 (\boldsymbol{\Omega}\times \boldsymbol{v})_{A} + \Big(\boldsymbol{\Omega} \times (\boldsymbol{\Omega} \times \boldsymbol{x})\Big)_{A} = 0 ~,
\label{Navierstokes}
\end{equation}
which is identical to \eqref{rotationNS1}.

The above equation \eqref{Navierstokes} is the Navier-Stokes equation (relative to the rotating frame) for a non relativistic, viscous fluid, with the last two terms being the inertial Coriolis and the centrifugal forces respectively, generated as a consequence of the relative fluid motion described in the rotating coordinate system. Thus the inertial effects of the Coriolis and the centrifugal force are codified inside the proposed metric dual. Again passing by we mention that the centrifugal force can be expressed as the gradient of a certain centrifugal potential,
\begin{eqnarray}
\Big(\boldsymbol{\Omega} \times (\boldsymbol{\Omega} \times \boldsymbol{x})\Big)_{A} = -\partial_A\Big(\frac{1}{2}\Big|\boldsymbol{\Omega}\times \boldsymbol{x}\Big|^2\Big) ~,
\end{eqnarray}
to identify an effective pressure $P_{\text{eff}}$ as ,
\begin{eqnarray}
P_{\textrm{eff}} = P - \frac{1}{2}|\boldsymbol{\Omega} \times \boldsymbol{x}|^2 ~,
\end{eqnarray}
thus obtaining \eqref{0.5} in the process.
\subsection{Incompressibility condition from expansion scalar}\label{ssection3}
Now in order to quantify the incompressibility condition of the fluid on the dual metric side we look at the deviation tensor of the geodesic congruence of the bulk velocity field and then project it on the $r = r_c$ hypersurface. We demand that this relevant quantity must vanish in analogy to the fact that on the fluid side incompressibility implies density perturbations being zero over the continuum macroscopic scales.
So we consider the tensor field $\nabla_b v_a$, which is the deviation of the geodesic fluid flow and then project this deviation tensor onto the $r=r_c$ timelike hypersurface. Basically we evaluate the term $\Theta = \gamma^{ab}(\nabla_b v_a)|_{r = r_c}$ which is the expansion scalar as seen on the timelike cut-off surface $r = r_c$. We shall see that the vanishing of this expansion scalar necessarily implies the incompressibility condition.

The corresponding projectors have been listed in \eqref{contraprojectors}. Hence expanding we have,
\begin{eqnarray}
\Theta &=& -\gamma^{\tau \tau} \Big(\Gamma^{\tau}_{\tau \tau}v_{\tau} + \Gamma^{r}_{\tau \tau}v_{r} + \Gamma^{A}_{\tau \tau} {v}_{A} \Big) - 2 \gamma^{\tau A}\Big(\Gamma^{\tau}_{\tau A}v_{\tau}+\Gamma^{r}_{\tau A}v_r + \Gamma^{A}_{\tau A}{v}_A\Big)\nonumber \\
&& +\gamma^{AB}\Big(\partial_A {v}_B -\Gamma^{\tau}_{AB}v_{\tau} - \Gamma^{r}_{AB}v_r + \Gamma^{D}_{AB}{v}_D\Big) ~.
\label{theta2}
\end{eqnarray}
Evaluation of the right hand side (RHS) of \eqref{theta2}, reveals that it vanishes at orders $\mathcal{O}(\epsilon^0)$, $\mathcal{O}(\epsilon^1)$ and $\mathcal{O}(\epsilon^3)$ (see Appendix \ref{appC} for details). The second order term $\mathcal{O}(\epsilon^2)$ implies (see Appendix \ref{appC}),
\begin{eqnarray}
\mathcal{O}(\epsilon^2): ~  \partial_A v^A \nonumber ~.
\end{eqnarray}
We impose the condition for the vanishing of the expansion scalar as evaluated on the $r = r_c$ cut-off hypersurface upto order $\mathcal{O}(\epsilon^3)$. This implies the incompressibility condition,
\begin{eqnarray}
 \partial_A {v}^A = 0 ~.
 \label{incompressibility}
 \end{eqnarray}
 This is perhaps physically intuitive as the expansion scalar contains information as to the expansion or the compression of the bulk geodesic velocity element and its vanishing simply translates to the incompressibility condition on the fluid side.

\section{FLUID DYNAMICS VIA PARALLEL TRANSPORT: Choice \rom{2}} \label{sssection1}
Again taking only the hydrodynamical scaling information from the fluid dynamical side \eqref{scalingssummary}, we construct another metric expanded order by order in terms of the hydrodynamic scaling parameter $\epsilon$. We propose a metric of the form,
\begin{eqnarray}
ds^2_{p+2}  = g_{ab}dx^a dx^b && =  -r d\tau^2 + 2 d\tau dr + dx_Adx^A \nonumber \\ 
&&-2\b_A f(r) d\tau dx^A -2 \b_A g(r) dr dx^A \nonumber \\ 
&&- \Big(\frac{2a_1}{r_c}\partial_A P_{\text{eff}}+{2 a_2} \partial^2v_A-\frac{4}{r_c}f(r)v^D \partial_D \b_A\Big)dx^A dr \nonumber \\
&& + \mathcal{O}(\epsilon^4)  ~.
\label{metrixproposal1}
\end{eqnarray}
The present metric again matches with the proposed metric in \cite{De:2019wok} (with $a_3$ being zero) provided $\b_A$ has been set to zero.
The first line is of order $\mathcal{O}(\epsilon^0)$ and is again the base Rindler metric, the second line is of order $\mathcal{O}(\epsilon^1)$  and the third line being of order $\mathcal{O}(\epsilon^3)$. Here we reiterate that $\b_A(\tau, x^B)$ scales as order $\mathcal{O}(\epsilon^1)$ and at the metric level no concrete connection can be made between $\b_A$ and the uniform angular velocity component $\Omega_A$ of the fluid side. However intuitively, $\b_A$ can be recognised along the lines of a ``rotation'' parameter of the spacetime. This has been done in analogy with the concept of frame dragging of inertial coordinates. If we assume for the moment that all the metric coefficients are independent of the $\tau$ coordinate (i.e the metric becomes stationary) and the $[A]$ angular coordinate, then frame dragging becomes a generic feature in such stationary spacetimes where $g_{\tau [A]} \neq 0$. The parentheses on $[A]$ imply selecting only one angular coordinate $A$, out of the $p$ ones. For such a metric, there will be two conserved quantities $p_{\tau}$ and $p_{[A]}$. The angular velocity of a particle dropped along the radial direction with zero conjugate momentum corresponding to the $[A]$ angular coordinate ($p_{[A]} = 0$) is $\frac{dx^{[A]}}{d\tau} = \frac{p^{[A]}}{p^{\tau}} = \omega(r,\bar{A})\neq 0$, where $\bar{A}$ refers to all the other angular coordinates without the single chosen $[A]$ coordinate. So a particle dropped radially, will acquire a non zero angular velocity. If we think passively about the particle being described in some local inertial frame where it is spatially at rest, then such inertial frames should be rotating with an angular velocity $\omega(r, \bar{A})$, and hence we say that inertial frames are dragged in this spacetime. It is in this sense that $\b_A$, which in general is a function of $(\tau,x^A)$ is identified as a parameter that describes the rotation of the above mentioned spacetime. The same conditions on $f(r_c)$ and $g(r_c)$ hold as did for metric choice $\rom{1}$ \eqref{metrixproposal}. Same condition holds for the location of the timelike cut-off slice $r = r_c$. However in \eqref{metrixproposal1}, $P_{\text{eff}}(\tau,x^A)$ is a modification of the field $P(\tau, x^A)$ as presented in \eqref{metrixproposal}. However we do demand that this modification is consistent with the scaling argument, i.e. $P_{\text{eff}}(\tau,x^A)$ scales as order $\mathcal{O}(\epsilon^2)$. We will at the end of the analysis decide what exact modification needs to be applied to $P_{\text{eff}}$ so that the duality between the manifold side and the fluid side is evident. The \textit{base bulk metric} here is 
\begin{eqnarray}
g^{(0)}_{ab}dx^{a}dx^{b} = -r d\tau^2 + 2 d\tau dr + dx_Adx^A -2\b_A f(r) d\tau dx^A -2\b_A g(r)drdx^A
\label{basemetrix1}
\end{eqnarray}
over which the \textit{perturbation}
\begin{eqnarray}
h^{(3)}_{ab} dx^{a}dx^{b} = - \Big(\frac{2a_1}{r_c}\partial_A P_{\text{eff}}+ {2a_2}\partial^2 v_A  -\frac{4}{r_c}f(r)v^D \partial_D \b_A \Big)dx^A dr
\label{perturbation31}
\end{eqnarray}
to the order $\epsilon^3$ has been applied. 
 We denote the perturbation as $h^{(3)}_{ab}$. As a result, the base bulk metric is curved, which can be checked by calculating the components of the Riemann curvature tensor $R^{a}_{~bcd}$.
 
 \subsection{Parallel transport of the velocity field} \label{sssection2}
The base bulk metric for the metric proposal \eqref{metrixproposal1} being written to order $\mathcal{O}(\epsilon)$, the unit normal on the $r=r_c$ timelike slice is evaluated to order first order in $\epsilon$. The computation yields,
\begin{equation}
n_a = \Big(0,\frac{1}{\sqrt{r_c}},\underset{\sim}{0}\Big) ~.
\end{equation}
The tilde underneath $0$, refers to all the angular components of the normal being zero. The components of the projection tensor (to order $\mathcal{O}(\epsilon)$) on the $r = r_c$ slice follow as 
\begin{eqnarray}
&& \g_{\tau \tau} = -r_c, \quad \quad \g_{\tau r} = 1, \quad \quad \g_{\tau A} = -\b_A f(r_c),\nonumber \\
&&\g_{rr} = -\frac{1}{r_c},\quad \quad \g_{rA} = -\b_A g(r_c) ,\quad \quad \g_{AB} = \delta_{AB} ~.
\label{projectors1}
\end{eqnarray} 
The contravariant components to order $\mathcal{O}(\epsilon)$ on raising via the metric are,
\begin{eqnarray}
&& \g^{\tau \tau} = -\frac{1}{r_c}, \quad \quad \g^{\tau r} = 0, \quad \quad \g^{\tau A} = -\frac{f(r_c)}{r_c}\delta^{AB}\b_{B}, \nonumber \\
&& \g^{rr} = 0, \quad \quad \g^{rA} = 0, \quad \quad \g^{AB} = \delta^{AB} ~.
\label{contraprojectors1}
\end{eqnarray}
Again for this proposed manifold defined via the metric \eqref{metrixproposal1}, we define a bulk velocity field as $v^a = (1,0,v^A)$. We follow exactly the same algorithm as done for the metric choice $\rom{1}$. The relevant information about the inverse metric components and the Christoffel symbols for the metric choice $\rom{2}$ have been listed in the Appendix (\ref{appA1}).

Setting the free index $c$ as $\tau$, for the metric \eqref{metrixproposal1}, we evaluate the L.H.S of \eqref{projectedautoparallel} order by order till $\mathcal{O}(\epsilon^3)$. We see that the L.H.S vanishes at $\mathcal{O}(\epsilon^0)$, $\mathcal{O}(\epsilon^1)$, $\mathcal{O}(\epsilon^2)$ and $\mathcal{O}(\epsilon^3)$, which has been detailed out in appendix (\ref{appB1}). Hence for the choice of the metric \eqref{metrixproposal1}, the projection of the parallel transport equation on the $r=r_c$ hypersurface is validated trivially.

Next, we look at the L.H.S of \eqref{projectedautoparallel} for the free index $c$ = $r$. Evaluating, we find again that it vanishes at $\mathcal{O}(\epsilon^0)$, $\mathcal{O}(\epsilon^1)$ and $\mathcal{O}(\epsilon^3)$. However at $\mathcal{O}(\epsilon^2)$ we have terms proportional to $\frac{f^2(r_c)}{r_c}+ 2 f(r_c)g(r_c)+r_{c} g^2(r_c)$. We are again presented with the same consistency condition on $f(r_c)$ and $g(r_c)$ for \eqref{projectedautoparallel} to be valid i.e $g(r_c) = -\frac{f(r_c)}{r_c}$, which has to be satisfied on the $r = r_c$ timelike slice.

The covariant components of the velocity field (lowered via the base bulk metric) to $\mathcal{O}(\epsilon^2)$ are,
\begin{eqnarray}
v_{\tau} = -r -f \b_A v^A ~, \quad v_r = 1-g\b_A v^A ~, \quad {v}_A = -f\b_A + \delta_{AB}v^B ~.
\end{eqnarray}
Evaluating the L.H.S of \eqref{projectedautoparallel} with the free index $c$ = $A$ reveals that it vanishes at $\mathcal{O}(\epsilon^0)$, $\mathcal{O}(\epsilon^1)$ and $\mathcal{O}(\epsilon^2)$. At order $\mathcal{O}(\epsilon^3)$, we have,
\begin{eqnarray}
\mathcal{O}(\epsilon^3): \gamma_{a A}(v^b \nabla_b v^a)|_{r=r_c} &=& \partial_{\tau} v_A + v^C\partial_C v_A \nonumber \\
&& + \frac{r_c}{2}\Big(\frac{a_1}{r_c}\partial_A P_{\text{eff}}+ a_2\partial^2v_A - \frac{2}{r_c}f(r_c)v^D\partial_D\beta_A\Big)\nonumber \\
&& + f(r_c)\Big(\partial_A \b_D -\partial_D \b_A \Big)v^D ~.
\label{projautoatA31}
\end{eqnarray}
For $\eqref{projectedautoparallel}$ to be valid, we demand that,
\begin{eqnarray}
\partial_{\tau} {v}_A+ {v}^C \partial_C {v}_A + \Big(\frac{a_1}{2}\partial_A P_{\text{eff}}+ \frac{a_2}{2} r_c \partial^2 {v}_A \Big) \nonumber \\
+ f(r_c)\Big(\partial_A \b_D - \partial_D \b_A \Big){v}^D &=& 0 ~. 
\label{asNavierStokes1}
\end{eqnarray}
So having arrived at \eqref{asNavierStokes1}, we want its correspondence to be made with \eqref{0.5}. As before we identify the field  $v_A(\tau, x^B)$ written at the metric level, to be indeed the velocity field of the fluid that needs to be described. In the same footing, we identify that the uniform angular velocity of the frame viewing the fluid system is related to the "rotation parameter" $\{\b_A\}$ on the gravity side via,
\begin{eqnarray}
(\partial_A \b_B - \partial_B \b_A){v}^B = -(\bs{\Omega} \times \bs{{v}})_A ~,
\end{eqnarray} as a result of which we can establish the analogy, 
\begin{eqnarray}
\boldsymbol{\Omega} = \bs{\nabla} \times \bs{\b} ~.
\label{Beta1}
\end{eqnarray}
 We constrain the functional form of $f(r)$ on the cutoff hypersurface as $f(r_c) = -2$ as a result of which $g(r_c) = \frac{2}{r_c}$.
Identifying the constants $a_i$'s in the proposed metric as \eqref{metrixproposal} as,
\begin{eqnarray}
a_1 = 2, \quad \quad a_2 = -2,
\label{constants1} 
\end{eqnarray}
and the kinematic viscosity term $\eta$ as $\eta = r_c$, \eqref{asNavierStokes} formalizes to,
\begin{eqnarray}
\partial_{\tau} {v}_A + {v}^C \partial_C {v}_A + \partial_A P_{\text{eff}} - \eta \partial^2 {v}+ 2(\levicivita{_{ABC}}\Omega^B {v}^C) = 0 ~.
\label{NavierStokes11}
\end{eqnarray}
So finally our correspondence with \eqref{NavierStokes} and \eqref{0.5} would be complete if we finalize the interpretation of $P_{\text{eff}}$.  We demand that,
\begin{eqnarray}
P_{\text{eff}} = P - \frac{1}{2} \Big[x_D x^D (\partial^A \b^B)(\partial_A \b_B -\partial_B \b_A) - x_A \levicivita{^{ABC}}(\partial_B \b_C) x^P \levicivita{_{PQR}}(\partial^Q \b^R)\Big] ~.
\label{Pressure_eff}
\end{eqnarray}
All the raising and lowering to the modification (in $P$) are done via the Euclidean metric.
The added term to $P$ i.e $-\frac{1}{2}  \Big[x_D x^D (\partial^A \b^B)(\partial_A \b_B -\partial_B \b_A) - x_A \epsilon^{ABC}(\partial_B \b_C) x^P \epsilon_{PQR}(\partial^Q \b^R)\Big]$ can be identified as $-\frac{1}{2}(\bs{\Omega}\times \bs{x})^2$, where $\bf{x}$ denotes the transverse coordinates on the $r=r_c$ hypersurface i.e it is the position vector of the velocity element on the  cut-off hypersurface and is in accordance to $\eqref{0.4}$.
The term $-\frac{1}{2}(\bs{\Omega}\times \bs{x})^2$ is identified as the centrifugal potential that needs to be added to the dynamical pressure to provide the effective pressure $P_{\text{eff}}$. We can clearly see here that $P_{\text{eff}}$ scales as order $\mathcal{O}(\epsilon^2)$.

The above equation \eqref{NavierStokes11} is the Navier-Stokes equation (w.r.t the rotating frame) for a non relativistic viscous fluid, with the last term being the inertial Coriolis force, $2({\bs{v}}\times \bs{\Omega})$, generated as a consequence of relative fluid velocity described in the rotating coordinate system. Thus  the inertial effects of the Coriolis and the centrifugal force are codified inside the proposed metric dual.

\subsection{Incompressibility condition from expansion scalar}\label{sssection3}
We go ahead as done previously to calculate the term $\Theta = \gamma^{ab}(\nabla_b v_a)|_{r = r_c}$ which is the expansion scalar as seen on the timelike cut-off surface $r = r_c$ for the manifold defined by \eqref{metrixproposal1}.The corresponding projectors have been listed in \eqref{contraprojectors1}. Hence expanding, we have,
\begin{eqnarray}
\Theta = \gamma^{ab}\Big(\partial_a v_b + r \Gamma^{\tau}_{ab} + f \b_D v^D \Gamma^{\tau}_{ab}- \Gamma^{r}_{ab} + g\b_D v^D \Gamma^{r}_{ab} - \Gamma^{D}_{ab} {v}_D + f \b_D \Gamma^D_{ab}\Big)|_{r =r_c} ~.
\label{theta11}
\end{eqnarray}
Neglecting terms of order greater than $\mathcal{O}(\epsilon^3)$, the above equation \eqref{theta11} simplifies as,
\begin{eqnarray}
\Theta &=& \frac{1}{r_c} \Big(\Gamma^{\tau}_{\tau \tau}v_{\tau} + \Gamma^{r}_{\tau \tau}v_{r} + \Gamma^{A}_{\tau \tau} {v}_{A} \Big)+\Big(2 \frac{f(r_c)}{r_c}\delta^{AB}\b_B\Big)\Big(\Gamma^{\tau}_{\tau A}v_{\tau}+\Gamma^{r}_{\tau A}v_r + \Gamma^{D}_{\tau A}{v}_D\Big)\nonumber \\
&& +\delta^{AB}\Big(\partial_A {v}_B -\Gamma^{\tau}_{AB}v_{\tau} - \Gamma^{r}_{AB}v_r + \Gamma^{D}_{AB}{v}_D\Big) ~.
\label{theta21}
\end{eqnarray}
Evaluation of the right hand side (RHS) of \eqref{theta2}, reveals that it vanishes at orders $\mathcal{O}(\epsilon^0)$, $\mathcal{O}(\epsilon^1)$ and $\mathcal{O}(\epsilon^3)$ (see Appendix \ref{appC} for a  derivation). The second order term $\mathcal{O}(\epsilon^2)$ implies (see Appendix \ref{appC1}),
\begin{eqnarray}
\mathcal{O}(\epsilon^2): ~ \sim \partial_A v^A  ~.
\end{eqnarray}
We impose the condition for the vanishing of the expansion scalar as evaluated on the $r = r_c$ cut-off hypersurface upto order $\mathcal{O}(\epsilon^3)$. This implies the incompressibility condition,
\begin{eqnarray}
 \partial_A {v}^A = 0 ~.
 \label{incompressibility}
 \end{eqnarray}
\section{CONSTRUCTION OF THE PROPOSED METRICS: A different interpretation}\label{sectionB}
  Now we propose to construct our two proposed metrics by a coordinate transformation on their respective base bulk metrics over which the perturbation at $\mathcal{O}(\epsilon^3)$ has been added to incorporate the forcing terms. About any event $\mathcal{P}$ in the manifold we employ the following coordinate transformations,
  \begin{align}
\tilde{x}^A &= x^A + \lambda \xi^{A \, (3)}_{b c} \, r \, , 
\nonumber
\\ 
\tilde{\tau} &= \tau \, , 
\nonumber
\\
\tilde{r} &= r \, ,  
\label{coordtransf}
\end{align}
where $\xi^{A \, (3)}_{b c} $ is an $3$ indexed component with $b$, $c$ taking values of either $\tau$ and $r$. 
We impose that 
\begin{equation}
\xi^{A \, (3)}_{b c} = \delta^{AB}\Big(\frac{a_1}{r_c}\partial_B P + a_2 \partial^2 v_B   -\frac{2}{r_c}f(r)v^D \partial_D \b_B \Big) 
\end{equation}
for the metric choice $\rom{1}$ \eqref{metrixproposal} and 
\begin{equation}
\xi^{A \, (3)}_{b c} = \delta^{AB}\Big(\frac{a_1}{r_c}\partial_B P_{\text{eff}} + a_2 \partial^2 v_B   -\frac{2}{r_c}f(r)v^D \partial_D \b_B \Big)
\end{equation}
for the metric choice $\rom{2}$ \eqref{metrixproposal1} such that the event $\mathcal{P}$ is taken to be the origin of both the coordinate systems and where $\lambda$ is simply a constant that shall be fixed in due course. So induction of these coordinate transformations on the initial base metric(s) which are genuinely curved do not hence change the overall structure of the spacetime. This is because the change in the Riemann curvature tensor $R^{a}_{~bcd}$ due to these coordinate transformations occur at $\mathcal{O}(\epsilon^4)$. As a result we have, for the first metric,
\begin{eqnarray}
d\tilde{x}^a &=& dx^a + \lambda\Big[\delta^{AB}\Big(\frac{a_1}{r_c}\partial_B P + a_2 \partial^2 v_B -\frac{2}{r_c}f(r)v^D \partial_D \b_B \Big)\Big] dr\nonumber \\
&& + \mathcal{O}(\geq \epsilon^4) ~.
\label{tranfx}
\end{eqnarray}
Similarly for the second choice of the metric we have,
\begin{eqnarray}
d\tilde{x}^a &=& dx^a + \lambda\Big[\delta^{AB}\Big(\frac{a_1}{r_c}\partial_B P_{\text{eff}} + a_2 \partial^2 v_B -\frac{2}{r_c}f(r)v^D \partial_D \b_B \Big)\Big] dr\nonumber \\
&& + \mathcal{O}(\geq \epsilon^4) ~.
\label{tranfx1}
\end{eqnarray}
Imposing the above transformations \eqref{coordtransf} on the base metric written to the second order \eqref{basemetrix}, we have,
\begin{eqnarray}
d\tilde{s}^2_{p+2} &=& -\tilde{r}d\tilde{\tau}^2 + 2d\tilde{\tau} d\tilde{r} + d\tilde{x}_Ad\tilde{x}^A -2 \b_A f(\tilde{r})d\tilde{\tau}d\tilde{x}^A - 2 \b_A g(\tilde{r}) d\tilde{r} d\tilde{x}^A \nonumber \\
&& + \Big(a_3(\tilde{x}^A \Omega_A \tilde{x}^B\Omega_B) + a_4(\delta^{AB}\Omega_A \Omega_B \delta_{CD}\tilde{x}^C\tilde{x}^D)\Big)d\tilde{\tau}^2 \nonumber \\
 && =   -r d\tau^2 + 2 d\tau dr + dx_Adx^A \nonumber \\ 
&&-2\b_A f(r) d\tau dx^A -2 \b_A g(r) dr dx^A \nonumber \\ 
&& +\Big(a_3(x^A\Omega_A x^{B}\Omega_B) + a_4(\delta^{AB}\Omega_A\Omega_B \delta_{CD}x^Cx^D)\Big)d\tau^2 \nonumber \\
&&+\lambda \Big(\frac{2a_1}{r_c}\partial_A P+{2 a_2} \partial^2v_A  -\frac{4}{r_c}f(r)v^D \partial_D \b_A\Big)dx^A dr\nonumber \\
&& + \mathcal{O}(\epsilon^4) ~.
\label{constructmetric}
\end{eqnarray}
Imposing the above transformations \eqref{coordtransf} on the base metric written to the first order \eqref{basemetrix1}, we have,
\begin{eqnarray}
d\tilde{s}^2_{p+2} &=& -\tilde{r}d\tilde{\tau}^2 + 2d\tilde{\tau} d\tilde{r} + d\tilde{x}_Ad\tilde{x}^A -2 \b_A f(\tilde{r})d\tilde{\tau}d\tilde{x}^A - 2 \b_A g(\tilde{r}) d\tilde{r} d\tilde{x}^A \nonumber \\
 && =   -r d\tau^2 + 2 d\tau dr + dx_Adx^A \nonumber \\ 
&&-2\b_A f(r) d\tau dx^A -2 \b_A g(r) dr dx^A \nonumber \\ 
&&+\lambda \Big(\frac{2a_1}{r_c}\partial_A P_{\text{eff}}+2 a_2 \partial^2v_A  -\frac{4}{r_c}f(r)v^D \partial_D \b_A\Big)dx^A dr\nonumber \\
&& + \mathcal{O}(\epsilon^4) ~.
\label{constructmetric}
\end{eqnarray}  
Setting the value of $\lambda$ $=$ $-1$, we obtain our proposed metrics upto $\mathcal{O}(\epsilon^3).$ So the effect of the forcing terms via the perturbation $h^{(3)}_{ab}$ at order $\mathcal{O}(\epsilon^3)$ present in the proposed metrics \eqref{metrixproposal} and \eqref{metrixproposal1} can be thought of the forces felt by observers moving in the spacetime defined by the base bulk metric i.e \eqref{basemetrix} and \eqref{basemetrix1} respectively, undergoing trajectories defined via \eqref{coordtransf}. Hence at least at the structural level, the original spacetimes $\eqref{metrixproposal}$ and \eqref{metrixproposal1} and the spacetimes \eqref{basemetrix} and \eqref{basemetrix1} respectively as observed by an observer following the trajectory \eqref{coordtransf} are inherently not different from each other at least to order $\mathcal{O}(\epsilon^3)$.

We mention that the base bulk metric (\ref{basemetrix}), for the proposed metric (\ref{metrixproposal}), is a genuinely curved manifold as evident from the calculation of the curvature components. The non zero components, upto order $\mathcal{O}(\epsilon^1)$, of $R^{a}_{~bcd}$ of the base metric upto order $\mathcal{O}(\epsilon^1)$ are evaluated to be 
\begin{eqnarray}
R^{r}_{~rAr} = \frac{f'' \b_A}{2} \quad \quad R^{r}_{~A\tau r} = -\frac{r f'' \b_A}{2} \quad \quad R^{r}_{~\tau Ar}  = -\frac{r f'' \b_A}{2}~;
 \nonumber 
 \\
R^{A}_{~r \tau r} = \frac{\delta^{AB} \b_B f''}{2} \quad \quad R^{\tau}_{~A \tau r} = -\frac{f'' \b_A}{2} \quad \quad R^{\tau}_{~\tau A r} = -\frac{f'' \b_A}{2}~.
\end{eqnarray}
Same  is also the case for the base metric (\ref{basemetrix1}). So, in general, even upto $\mathcal{O}(\epsilon^1)$, the Riemann curvature tensor has non-vanishing components. Therefore both the base bulk metrics (\ref{basemetrix}) and (\ref{basemetrix1}), unlike that in \cite{De:2019wok}, are curved. This is a very crucial difference between the earlier proposal and the present one.  In this regard, it is worth mentioning that on the fluid side a simple set of coordinate transformations allow us to transform between the NS equation written in the inertial coordinates and the uniformly rotating non inertial coordinates. Therefore, one can expect that similar argument can be applied to construct the metric for rotating case. So the proposed metric in \cite{De:2019wok} may provide that for the rotating situation by just a simple coordinate transformation. In that respect, the base metric should be flat in both the situations. But unfortunately, that is not the case. As we mentioned above that the base metrics in the present discussion are curved in nature, so coordinate transformation can not connect them with the metric presented in \cite{De:2019wok}. This clearly shows that on the gravity side, the idea is not so simple. Hence the obtention of gravity dual of fluid equation in rotating frame needs a special attention.  In this analysis we have precisely done the same. In addition, what we found is that the parameter which is connected to the intrinsic rotation of the spacetime (i.e. $\beta_A$), provides the rotational effect in the fluid side. This shows a clear correspondence between the parameters on both sides and interestingly non-vanishing of $\beta_A$ guarantees the non-flatness of the base metrics. Hence we feel that the analysis, done here, adds a non-trivial contribution in the subject of fluid-gravity correspondence.  

\section{DISCUSSIONS AND OUTLOOK}\label{sectionC}
We summarize our calculations as follows. We have  proposed two bulk metrics in $(p+2)$ dimensions such that the base bulk metrics that act as the background are genuinely curved manifolds. To the zeroth order in the hydrodynamic parameter $\epsilon$, the background is essentially the flat Rindler spacetime for both the spacetimes.
Onto this background is added the perturbation $h^{(3)}_{ab}$ at $\mathcal{O}(\epsilon^3)$ which contains the information about both the pressure $P(\tau, x^A)$ and the velocity $v_A(\tau, x^B)$ fields. The perturbation contains all the ``forcing'' terms i.e the forces due to pressure gradients and the viscous effect. 
We then choose a bulk velocity vector field contained in this bulk spacetime defined as $v^a = (1,0,v^A)$ . Our basic formalism involves the fact that this appropriately chosen bulk velocity has no component of its acceleration on the $r=r_c$ timelike hypersurface as evident from from \eqref{projectedautoparallel} which we imply as the projection of the parallel transport equation. We look at the projection of this parallel transport equation on a specific safely chosen $r =r_c$ timelike hypersurface such that the location of the slice is away from the horizon (at $r =0$) and $r \rightarrow \infty$ as shown in \eqref{projectedautoparallel}. Demanding the consistency of the projection of this parallel transport equation onto the $r=r_c$ hypersurface to be evident till order $\mathcal{O}(\epsilon^3)$ (since the metric has been constructed to $\mathcal{O}(\epsilon^3)$), requires that the incompressible NS equation (relative to the rotating frame) be satisfied at $\mathcal{O}(\epsilon^3)$ along with the consistency conditions on $f(r_c)$ and $g(r_c)$ being generated at $\mathcal{O}(\epsilon^2)$. The relevant centrifugal force is generated at order $\mathcal{O}(\epsilon^3)$ (via the parallel transport formalism) in the computation of the Christoffel connection component $\Gamma^{A}_{\tau \tau}$. Similarly, the Coriolis force is generated via the component $\Gamma^{A}_{\tau B}$. We then go ahead to show that demanding a vanishing expansion scalar for the bulk velocity field defined in the spacetime (as seen from the $r=r_c$ slice) leads to the incompressibility condition for the viscous fluid as observed from the rotating frame. Finally we come up with the systematics of constructing the proposed metrics \eqref{metrixproposal} and \eqref{metrixproposal1}via a coordinate transformation on the base bulk metrics.

In the previous work \cite{De:2019wok}, the authors had proposed a metric dual to the incompressible nonrelativistic NS equation that contained (within the metric) all information about the forcing terms i.e the forces of pressure gradient and the viscous effects. In that work, the analogy was of the correspondence between the dynamics of a viscous, non relativistic incompressible fluid in the Minkowski flat spacetime (described by the NS equation) with the dynamics of a free fluid (described by the parallel transport equation) residing in a manifold given by a proposed choice of metric. The present work is an extension based on the formalism proposed in the earlier paper \cite{De:2019wok}. Here however the two proposed metrics are distinctly different along with the fact that the base bulk metric(s) are no longer flat \eqref{basemetrix} and \eqref{basemetrix1}. Similarly there are additional terms that occur in the perturbation at order $\mathcal{O}(\epsilon^3)$ for the two proposed metrics. The present two metrics contain all the effects of the forcing terms along with bringing in the centrifugal and the Coriolis forces. This is because of the analogy that needs to be setup between  the dynamics of an incompressible viscous fluid (as viewed from a rotating frame) with that of a free bulk fluid (being parallel transported along its own geodesic integral curves) residing in a manifold given by the proposed two metrics \eqref{metrixproposal} and \eqref{metrixproposal1}. When the fluid is to be described in a rotating coordinate system, then the metric is no longer diagonal in structure but rather involves cross terms between the temporal and spatial coordinates. On the dual metric side the effect of rotation is induced by the "rotation parameter" $\{\beta_A\}$ which causes the base bulk metric to be no longer flat. Projection of the parallel transport equation of the appropriately chosen bulk fluid velocity field on the $r =r_c$ timelike slice gives rise to the required dynamics on the fluid side with the additional Coriolis and centrifugal forces being generated. The forcing terms of pressure and viscosity are encoded in $h^{(3)}_{ab}$ of the proposed metrics.  

We can hence gain some new perspectives into the dynamical structure of the incompressible NS fluid equations relative to a rotating coordinate system. Rewriting them in the $\boldsymbol{F} = m \boldsymbol{a}$ form,
\begin{eqnarray}
\partial_{\tau}{v_A} + {v^D}\partial_D {v_A} = -\partial_A P + \eta \partial^2 v_A - \Big(\boldsymbol{\Omega}\times (\boldsymbol{\Omega} \times \boldsymbol{x})\Big)_A - 2 \Big(\boldsymbol{\Omega}\times {\boldsymbol{v}}\Big)_A ~,
\label{newtonian_form}
\end{eqnarray}
where the L.H.S is the total derivative for the velocity of the fluid element relative to the rotating system. The R.H.S contains the regular forcing terms due to pressure and viscosity along with the additional inertial centrifugal and Coriolis forces due to the system being described in a rotating frame. As is evident from \eqref{projautoatA3}, these forcing terms along with the inertial forces essentially arise from the evaluation of the relevant Christoffel symbols for the metric \eqref{metrixproposal} and \eqref{metrixproposal1}. In the previous paper by the authors \cite{De:2019wok}, all the forcing terms were built inside the perturbation $h^{(3)}_{ab}$. Here, in this paper for the metric choice $\rom{1}$ the pressure and the viscous forces are generated from $h^{(3)}_{ab}$. The centrifugal force is generated through the metric component $g^{(2)}_{\tau \tau}$. The Coriolis force is generated from the rotation parameter $\beta(\tau,x^A)$ that shows its effect in the metric at order $\mathcal{O}(\epsilon)$. So if we were to "switch-off" this perturbation of the fields by putting $P(\tau,x^A) = 0$ and $v_A(\tau,x^A) = 0$, then we would have a "forcing-free" fluid as described relative to a rotating coordinate system i.e
\begin{eqnarray}
\partial_{\tau}{v_A} + v^D\partial_D {v_A} = \Big(\boldsymbol{\Omega}\times (\boldsymbol{\Omega} \times \boldsymbol{x})\Big)_A - 2 \Big(\boldsymbol{\Omega}\times {\boldsymbol{v}}\Big)_A ~.
\end{eqnarray}  
Similarly, for the metric choice $\rom{2}$, the dynamical pressure (along with the centrifugal forces) and the viscous forces are incorporated in $h^{(3)}_{ab}$ of the metric. The Coriolis force is again generated from the rotation parameter $\beta(\tau,x^A)$ at order $\mathcal{O}(\epsilon)$ in the metric. Hence the correspondence is that of a viscous incompressible fluid residing in a flat spacetime (however not Minkowski in structure) being dynamically equivalent to the geodesic flow of a free fluid (appropriately defined) in a curved manifold defined via \eqref{metrixproposal} or \eqref{metrixproposal1}. This actually in a sense parallels the interpretation where the dynamics of a particle interacting in a static gravitational field is locally indistinguishable from an equivalent accelerated frame, which has been expounded in \cite{Paddybook} (refer to Sec. $3.3$ of this book).

Hence we have observed that the present two metrics in this paper account for all the forcing as well as the inertial terms of the NS equation. Hence the behaviour of a free fluid in the proposed metrics can be considered as an \textit{equivalent theory of non-relativistic viscous fluid dynamics relative to a uniformly rotating frame}. As a result a strong parallel can be brought in between this present analysis and the equivalence principle of gravity where an appropriate accelerated frame can locally mimic gravity. Thus the duality presented here can be a dictionary between calculations on both the sides. Any calculation that may be hard to extract on the fluid side (of the incompressible viscous flow in the rotating frame) of the NS equation can be reflected by calculation on a free fluid in the proposed metric spacetime (which incorporates both the effects of the forcing and the inertial effects) or vice versa. As a suggestive interpretation these metrics can be thought of as a complete \textit{geometrical description} of the NS equation w.r.t the rotating frame.

We have as a consequence of the projection of the parallel transport equation on the induced timelike hypersurface, the NS equation. In fact any corresponding alteration to the NS equation (relative to the rotating frame) at one's own convenience can be supplemented with corresponding term(s) on the metric side and the present algorithm would yield for the person his/her choice of equation. That is, one can add corrections to the NS equation and construct the dual metric easily. So the question naturally arises as to why we are fixated on the NS equation written in the rotating frame in our analysis. This is simply because we are in the purview of the non-relativistic regime and the hydrodynamic limit($\epsilon \rightarrow 0$). In this regime, the incompressible NS equation is universally the hydrodynamic limit to essentially any fluid system. Any corrections that can be thought of either coming from kinetic theory or from the theory of strongly coupled fluids will necessarily get scaled away in this limit. We have hence constructed/formulated two metric duals to a viscous fluid system viewed from a uniformly rotating frame in this particular limit and singled out the NS equation as a consequence of parallel transport on these curved manifolds. 

We again stress at this point reminding that in the present analysis the Einstein's equation have not been used to bring the correspondence between fluid dynamics and gravitational dynamics. In most of the earlier interpretations of fluid/gravity correspondence \citep{Bhattacharyya:2008kq, Bhattacharyya:2008jc, Bhattacharyya:2008ji, Policastro:2002se, Bredberg:2011jq, Huang:2011he, De:2018zxo, Zhang:2012uy, Chirco:2011ex, Bai:2012ci, Cai:2012mg, Zou:2013ix, Hu:2013lua, Cai:2011xv, Huang:2011kj, Anninos:2011zn, Ling:2013kua, Eling:2012ni, Berkeley:2012kz, Lysov:2017cmc, Wu:2013aov} the Einstein's equation played a pivotal role. In the holographic approach to fluid/gravity correspondence, where the Einstein's equation was interpreted as the NS equation on a timelike slice is actually one of the possible ways to bring about this correspondence. There is absolutely no requirement that the Einstein's equation have to be used to connect the dynamics of both these two sectors. We have shown in our analysis, this correspondence using the parallel transport equation as our guiding principle. Hence, our approach can be designated as an \textit{off-shell} approach to fluid/gravity correspondence. 
The authors of \cite{Pinzani-Fokeeva:2014cka} have generalized the result of \cite{Bredberg:2011jq} by describing the dynamics of the fluid on the cut-off surface. Here the authors have considered a general curved static metric rather than the flat Rindler metric, but have kept the induced surface $r=r_c$ flat. By performing a set of scale transformations and  Lorentz transformations, they obtained the seed metric for the relativistic fluid dual on the cut-off surface. However, our approach is based on projection of the parallel transport equation on the $r=r_c$ surface. Moreover in our case, the $r=r_c$  surface in not flat because of the introduction of the rotation parameters $\b_A$. Our work differs from their cut-off surface based approach again in the sense that our's is an \textit{off-shell} analysis. However we have not been able to consider the duality of the metric with a relativistic dual fluid.

We now mention some points which we perceive are the apparent benefits of such an \textit{off-shell} approach to fluid/gravity duality and thereby may provide possible future aspects.

\noindent
$\bullet$ First is the possibility of constructing an action for the fluid system from such a set-up. The idea is as follows. The dynamics of the fluid are here encoded in the manifold properties of the considered spacetime(s) which are a priori not demanded to be solutions of the Einstein's equations. Using the duality between the fluid side and the manifold side via the projection of the parallel transport equations, the fluid system can be considered to be a collection of particles that are parallel transported on the $r=r_c$ hypersurface of the proposed spacetime(s). For such a fluid particle the action can be written as $\mathcal{A} = \int \sqrt{-g_{ab} u^a u^b} d\lambda$, where $u^a = dx^a/d\lambda$ is the velocity of the fluid particle and $g_{ab}$ is the metric of  our proposed spacetimes. However we do need to find a relation between $u^a$ and $v^A$ and then the action of the fluid particle can be expressed in terms of $v^A$. Extremizing such an action written for such a collection of fluid particles parallel transported on the $r= r_c$ slice in the proposed spacetimes might yield for us the required NS equation. In this way an action principle of NS equation may be constructed.

\noindent
$\bullet$ As a classical correspondence, we have shown in our analysis the map between classical fluid dynamical equations and the equations of the projected parallel transport of an appropriately defined bulk fluid velocity on the $r=r_c $ timelike slice of the given spacetime(s). However there is a difficulty that arises when we try to have a quantum theory on both the sides. In the on-shell approaches to fluid/gravity correspondence, we do have a quantum theory of the fluid. However a quantum theory of gravity is as of yet in progress. Our \textit{off-shell} approach may help to bypass this problem as we have not used the Einstein's equations. On the fluid side, we have a many-body interacting theory of the fluid living in Minkowski spacetime. The analogous quantum theory on the manifold side in our approach is that of quantizing a collection of ``free'' or parallel transported fluid particles in the background of the $r=r_c$ timelike slice of the proposed spacetime(s). This is basically semi-classical gravity where the background remains classical and we quantize matter fields in this background. Hence the quantum theories on both the sides can be related, bypassing the issues pertaining to quantum gravity. 

\noindent
$\bullet$ Another aspect that this \textit{off-shell} approach may help is in uncovering is the  microscopic degrees of freedom of the thermodynamic aspects of gravity. Our present formalism of fluid/gravity duality implies that all the degrees of freedom (dof) of the fluid are possibly encoded in the manifold properties of the concerned proposed spacetime(s). Therefore a working knowledge about the microstructure of the fluid and hence its thermodynamics may help to understand those of gravity. Moreover, it is now well known that one can associate quantities like entropy density and temperature with an arbitrary null hypersurface \cite{Parattu:2013gwa}. The null hypersurface does not need to be a solution to the Einstein's equations. However a proper origin of these quantities from the underlying microscopic structure of the spacetime is missing. Our \textit{off-shell} approach to fluid/gravity duality may help to elucidate the microscopic dof of the gravity side since we have a mapping between the fluid and the gravity side.  Previous correspondences between fluid and gravity, like the cut-off surface approach have been explicitly via the equations of motion. Hence we understand that in such approaches the dof of the fluid are encoded in the spacetimes that are required to be solutions to the Einstein's equations. But as the thermodynamics entities can be assigned with arbitrary null hypersurface and are related to the manifold properties which as such do not require to be solutions of the Einstein's equations,  it may be much more relevant to have an \textit{off-shell} duality approach to identify the dof of the manifold. In this regard our present approach may provide some light towards the thermodynamical origins of gravity.

In \cite{De:2019wok}, we had added upon the base flat Rindler metric a correction at order $\mathcal{O}(\epsilon^3)$ that incorporated all the forcing terms such that the demand of the projection of the parallel transport equation on the $r=r_c$ slice yielded for us the NS equation at $\mathcal{O}(\epsilon^3)$. The demand for the projection of the expansion scalar on the $r=r_c$ slice to vanish gave us the incompressibility condition at order $\mathcal{O}(\epsilon^2)$. It would definitely be interesting to construct the metric to the $n$th order as has been carried out in \cite{Compere:2011} to all orders such that we also, in our case retrieve the NS equation and the incompressibility condition along with the necessary corrections in the higher orders. The basis behind the construction of the metric in \cite{Compere:2011} to the $n$th order is as follows. Using the parallel of the hydrodynamic expansion of the fluid, the authors in \cite{Compere:2011} construct the bulk expansion of the metric via a gradient expansion to all orders in $\epsilon$. Demanding Ricci flatness to all orders (which is a partial differential equation), the gradient expansion imposes a hierarchy between the derivatives which converts the partial differential equation into a series of coupled ordinary differential equations. Assuming that the metric has been written to order $\epsilon^{n-1}$, the authors add a new term $g_{ab}^{(n)}$ at order $\epsilon^n$ as a result of which the Ricci tensor at order $\epsilon^n$ is $R_{ab}^{(n)} = \delta R_{ab}^{(n)} + \hat{R}_{ab}^{(n)}$, where $ \hat{R}_{ab}^{(n)}$ is the non linear contribution from the metric written till order $\epsilon^{n-1}$ and $\delta R_{ab}^{(n)}$ is the linearized contribution at order $\epsilon^{n}$ that contains only the $r$ derivatives. Demanding $R_{ab}^{(n)} = \delta R_{ab}^{(n)} + \hat{R}_{ab}^{(n)} = 0$, the Ricci flatness condition is then integrated to find the corrections to the metric at order $\epsilon^n$ to the preexisting one written till order $\epsilon^{n-1}$. There are integrability conditions that need to be satisfied for these equations to be integrated. It turns out that the conservation of the Brown-York stress tensor on the $r=r_c$ slice at order $\epsilon^n$ ensures the validity of the integrability conditions. Mapping it onto the dual fluid side, this conservation yields the NS equation along with its corrections for all odd orders, while the conservation yields the incompressibility condition along with its corrections for all even orders.

Now we step back to understand if it is possible in our present scheme to have a bulk construction of the metric to all orders in $\epsilon$, such that the projection of the parallel transport equation of an appropriately defined velocity field in this (bulk constructed to all orders) metric spacetime gives the NS equation along with its corrections. 
	However we have to build the metric with certain restrictions. Identification of those is non-trivial. Till now we have not been able to find those. Since we are aiming to present an \textit{off-shell} description, it is not desirable to use any information from the Einstein's equations. One such natural way of constructing the higher order metric of \cite{De:2019wok} is via coordinate transformations. We have seen that the metric written to the third order in $\epsilon$ in \cite{De:2019wok} has been generated by coordinate transformations on the base flat Rindler metric. Our goal is to construct the higher order terms in the metric via such diffeomorphism transformations. We calculate $\delta g_{ab}^{(4)}$, which is the fourth order contribution to the metric due to the application of the coordinate transformation $x^a \rightarrow x^a + \xi^{(4)a}(x)$ on the $\epsilon^3$ order seed metric. Hence one needs to find the diffeomorphism vector $\xi^{(4)a}(x)$ by taking the variation as Lie one, i.e. $\delta g_{ab}^{(4)} = \pounds_{\xi^{(4)}} g_{ab}$, under certain conditions on the choice of the metric coefficients, which we shall state later.
As a result of this we have the following equations,
	\begin{eqnarray}\label{lieequations}
		&\delta g_{rr}^{(4)} &= 2 \partial_r \xi^{(4)\tau}\nonumber \\ 
		&\delta g_{r \tau}^{(4)} &= -r \partial_{r} \xi^{(4) \tau} + \partial_r \xi^{(4)r}\nonumber \\
		&\delta g_{rA}^{(4)} &= \partial_r \xi_{A}^{(4)}\nonumber \\
		&\delta g_{\tau \tau}^{(4)}& = -\xi^{(4)r}\nonumber \\
		&\delta g_{\tau A}^{(4)} &= 0\nonumber \\
		&\delta g_{AB}^{(4)} &= 0~.
	\end{eqnarray}
	Because of the fact that we do not have boundary conditions prescribed for the components $\delta g_{r a}^{(4)}$ on the $r=r_c$ timelike slice (as for $r=$ constant surface $dr=0$ for any value of $g_{ra}$), we can set the constraint that $\delta g_{r a}^{(4)} = 0$. In fact this constraint needs to be extended to all orders such that $\delta g_{r a}^{(n)} = 0$ and hence the $g_{r a}$ components are actually those of the metric components generated to the third order in $\epsilon$. That is to all orders in the metric construction we have the imposition that,
	\begin{equation}
		g_{rr} = 0; \quad \quad g_{r \tau} = 1; \quad \quad g_{r_A} = \left(\frac{a_{1}}{r_{c}} \partial_{A} P+a_{2} \partial^{2} v_{A}+\frac{a_{3}}{r_{c}} \partial_{A} v^{2}\right)~.
	\end{equation}
	Now the above set of equations (\ref{lieequations}) need to be solved  for $\xi^{(4)a}$ with the boundary conditions that $\delta g_{\tau \tau}^{(4)}$, $\delta g_{\tau A}^{(4)}$ and $\delta g_{A B}^{(4)}$ all vanish on the $r =r_c$ timelike slice. This fixes the integration constants upon the solution of the equation (\ref{lieequations}). Now once we have solved for $\xi^{(4)a}$, we can apply the coordinate transformation $x^a \rightarrow x^a + \xi^{(4)a}(x)$ on the metric written to the third order in $\epsilon$ to generate the fourth order correction to the metric. After this we apply our machinery of the projection of the parallel transport equation of the bulk velocity field on the $r=r_c$ slice and generate the NS equation. This process of constructing the metric can be then be iterated to all orders in $\epsilon$ via this idea of coordinate transformations on the previous order metric. Equation of the projection of the parallel transport equation in the metric written to all orders in $\epsilon$ then yields the NS equations along with their corrections to the higher orders.

Another such possibility of construction of the metric to higher orders may be via $\nabla_ag_{bc}=0$ at all orders as we are doing psuedo Riemannian geometry. The idea is similar to that presented in \cite{Compere:2011}. We shall add a correction to the metric $\delta g_{ab}^{(n)}$ to the preconstructed metric written till order $\epsilon^{n-1}$. Since the condition is satisfied till our $\epsilon^{n-1}$ order metric, we choose the corrections at $\epsilon^n$ order such that the following is satisfied:
\begin{eqnarray}
\Big(\widehat{\nabla_{a}g_{bc}\Big)}^{(n)} + \nabla_a \Big(\delta g_{bc}^{(n)}\Big) = 0 ~,
\end{eqnarray}
where $\Big(\widehat{\nabla_{a}g_{bc}}\Big)^{(n)}$ is the $n$th order contribution due to the metric written till order $\epsilon^{n-1}$ and $\nabla_a \Big(\delta g_{bc}^{(n)}\Big)$ is the contribution due to $\delta g_{ab}^{(n)}$. This yields for us,
\begin{equation}
\partial_r \Big(\delta g_{bc}^{(n)}\Big) = \Big(\widehat{\nabla_{a}g_{bc}\Big)}^{(n)} + \Gamma_{ba}^{i(0)} \Big(\delta g_{ic}^{(n)}\Big) + \Gamma_{ca}^{i(0)}\Big(\delta g_{bi}^{(n)}\Big) ~,
\end{equation}
which can be integrated to find the corrections to the metric. Next with the imposition of the projection of the parallel transport equation of $v^a$ on the $r=r_c$ slice the higher order corrections to NS equation can be found out. Then from $\theta$, the corrections to incompressibility condition can also be evaluated.  However we are in the process of a progress in this direction and a thorough investigation is required. Hence nothing can be concretely stated right now. The work is in progress and will be reported in due time. In addition we mention that this \textit{off-shell} construction of the metric to all order for the initial choice the metric (\ref{metrixproposal}) and (\ref{metrixproposal1}) will be more non-trivial since the base metrics in both the cases are not Rindler flat, but rather genuinely curved spacetimes. As of now, we do not know which of the two procedures is correct, but are looking into this. We certainly aim to report our investigations pertaining to these issues in the near future.

Finally, we mention that the above formalism can be extended to yield the Damour NS equation for a viscous fluid relative to a rotating frame if we modify the proposed metric by adding a certain term to $h^{(3)}_{ab}$ for both the metrics. The term at order $\mathcal{O}(\epsilon^3)$ that does the job is $-\frac{2}{r_c}\partial_A v^2$. Following the exact same formalism outlined here, the rotating Damour NS equation has the form,
\begin{eqnarray}
\partial_{\tau} {v}_A + {v}^C \partial_C {v}_A + \frac{1}{2} \partial_A {v}^2+  \partial_A P_{\text{eff}} - \eta \partial^2 {v}+ 2(\levicivita{_{ABC}}\Omega^B {v}^C) = 0 ~.
\label{DamourNavierStokes}
\end{eqnarray}

Overall we hope that the present discussion will shed more light to the subject of fluid-gravity correspondence as it provides a new way of investigation.  


\section*{Appendices}
\appendix
\section{The Inverse Metric and the Christoffel symbols of metric choice \rom{1}}\label{appA}
We evaluate the inverse metric corresponding to \eqref{metrixproposal} as a perturbation series over the flat metric $g_{ab}^{(\text{flat})}dx^a dx^b$ =  $-r d\tau^2 + 2 d\tau dr + dx_Adx^A$ with the perturbation identified as $H_{ab} dx^a dx^b$ = $-2\b_A f(r) d\tau dx^A -2 \b_A g(r) dr dx^A +\Big\{a_3(x^A\Omega_A x^{B}\Omega_B) + a_4(\delta^{AB}\Omega_A\Omega_B \delta_{CD}x^Cx^D)\Big\}d\tau^2\\-\Big(\frac{2a_1}{r_c}\partial_A P +{2 a_2} \partial^2v_A -\frac{4}{r_c}f(r)v^D \partial_D \b_A\Big)dx^A dr$. The perturbation contains terms of $\mathcal{O}(\epsilon)$, $\mathcal{O}(\epsilon^2)$  and $\mathcal{O}(\epsilon^3)$. The inverse metric written as a perturbation series over the flat Rindler base metric,
\begin{equation}
g^{ab} = g^{ab(0)} - H^{ab} + H^{ac}H_{c}^{b} - H^{ai}H_{i}^{k}H_{k}^{b} +\mathcal{O}(H^4)
\label{perturbationinverse}
\end{equation} 
where all the raising has been performed via the flat base metric. We list below the inverse metric upto order $\mathcal{O}(\epsilon^3)$.
\begin{eqnarray}
g^{\tau \tau} &=& g^2 \delta^{AB}\b_A \b_B +\mathcal{O}(\epsilon^4) \\
g^{\tau r} &=& 1 + (rg^2 + fg)\delta^{AB}\b_A \b_B +\mathcal{O}(\epsilon^4) \\
g^{\tau A} &=& g \delta^{AB} \b_B + \delta^{AB}\Big(\frac{a_1}{r_c}\partial_B P+ a_2\partial^2 v_B -\frac{2}{r_c}f(r)v^D \partial_D \b_B \Big) \nonumber \\ 
&& + (2fg^2 + r g^3)\delta^{AB} \delta^{CD}\b_B \b_C \b_D +\mathcal{O}(\epsilon^5) \\
g^{rr} &=& r + (f^2 + r^2 g^2 +2 rfg)\delta^{AB}\b_A \b_B \nonumber \\
&& - a_3(x^A\Omega_A x^{B}\Omega_B)-  a_4(\delta^{AB}\Omega_A\Omega_B \delta_{CD}x^Cx^D) +\mathcal{O}(\epsilon^4)\\
g^{rA} &=& (f+ rg)\delta^{AB}\b_B + r \delta^{AB}\Big(\frac{a_1}{r_c}\partial_B P +{a_2}\partial^2 v_B -\frac{2}{r_c}f(r)v^D \partial_D \b_B \Big) \nonumber \\
&& +(2f^2 g + 3rfg^2 +r^2g^3)\delta^{AB}\delta^{CD} \b_B \b_C \b_D \nonumber \\
&& -g \delta^{AB}\b_{B}\Big[a_3(x^P\Omega_P x^{Q}\Omega_Q) + a_4(\delta^{CD}\Omega_C\Omega_D \delta_{PQ}x^Px^Q)\Big] + \mathcal{O}(\epsilon^5)\\
g^{AB} &=& \delta^{AB} + (rg^2 + 2fg)\delta^{AC}\delta^{BD}\b_C \b_D + \mathcal{O}(\epsilon^4)~.
\end{eqnarray}
In the same vein, we calculate the Christoffel symbols upto order $\mathcal{O}(\epsilon^3)$ with them being, 

\begin{eqnarray}
\Gamma^{\tau}_{\tau \tau} &=& \frac{1}{2} + \frac{1}{2}(rg^2 + fg)\delta^{AB}\b_A \b_B + \mathcal{O}(\epsilon^4)\\
\Gamma^{\tau}_{\tau r} &=& -\frac{1}{2}(g^2 + f'g)\delta^{AB} \b_A \b_B  +\mathcal{O}(\epsilon^4)\\
\Gamma^{\tau}_{\tau A} &=& \frac{1}{2} f'\b_A + \frac{1}{2}(rg^2f' + f'fg)\b_A \delta^{CD}\b_C \b_D - \frac{1}{2}g \partial_{\tau}\b_A \nonumber \\
&& + \frac{1}{2} fg \delta^{BC}\b_C (\partial_{B}\b_A - \partial_A \b_B) +\mathcal{O}(\epsilon^5) \\
\Gamma^{\tau}_{r r} &=& -g'g \delta^{AB}\b_A \b_B +\mathcal{O}(\epsilon^4) \\
\Gamma^{\tau}_{r A} &=& -\frac{1}{2} f'g^2 \b_A \delta^{CD}\b_C \b_D + \frac{1}{2}g^2\delta^{BC}\b_C(\partial_B \b_A - \partial_A \b_B) + \mathcal{O}(\epsilon^5) \\
\Gamma^{\tau}_{AB} &=& -\frac{1}{2}g(\partial_A \b_B +\partial_B\b_A) +\mathcal{O}(\epsilon^4) \\
\Gamma^{r}_{\tau \tau} &=& \frac{r}{2} + \frac{1}{2}(f^2 + r^2g^2 + 2rfg)\delta^{AB}\b_A \b_B \nonumber \\
&& -\frac{1}{2}\Big[a_3(x^A\Omega_A x^{B}\Omega_B) + a_4(\delta^{AB}\Omega_A\Omega_B \delta_{CD}x^Cx^D)\Big]+\mathcal{O}(\epsilon^4) \\
\Gamma^{r}_{\tau r} &=& -\frac{1}{2} -\frac{1}{2}(f'f + rgf' +rg^2 +fg)\delta^{AB}\b_A \b_B +\mathcal{O}(\epsilon^4) \\
\Gamma^{r}_{\tau A} &=& \frac{1}{2}r f' \b_A + \frac{1}{2}(f^2f' + r^2 g^2 f' + 2 rf'fg)\b_A \delta^{BC}\b_B\b_C \nonumber \\
&& -\frac{r}{2} g \partial_{\tau}\b_A + \frac{1}{2}(f^2 + rfg)\delta^{BC}\b_C(\partial_{B}\b_A -\partial_A \b_B) \nonumber \\
&& + a_3 \Omega_A(\Omega_D x^D) + a_4 (\delta^{PQ}\Omega_P \Omega_Q \delta_{AB}x^B) \nonumber \\
&& -\frac{1}{2}  f' \b_A \Big\{a_3(x^C\Omega_C x^{D}\Omega_D) + a_4(\delta^{CD}\Omega_C\Omega_D \delta_{PQ}x^Px^Q)\Big\}+ \mathcal{O}(\epsilon^5)\\
\Gamma^{r}_{rr} &=& -(f g' + rgg')\delta^{AB}\b_A \b_B +\mathcal{O}(\epsilon^4)\\
\Gamma^{r}_{r A} &=& -\frac{1}{2}f' \b_A -\frac{1}{2}(f'rg^2 + f'fg)\b_A \delta^{BC}\b_B \b_C \nonumber \\
&& + \frac{1}{2}g \partial_{\tau} \b_A + \frac{1}{2}(fg +rg^2)\delta^{BC}\b_C(\partial_B \b_A -\partial_A \b_B) +\mathcal{O}(\epsilon^5) \\
\Gamma^{r}_{AB} &=& -\frac{1}{2}(f+rg)(\partial_A \b_B + \partial_B \b_A) +\mathcal{O}(\epsilon^4) \\
\Gamma^{A}_{\tau \tau} &=& \frac{1}{2}(f + rg)\delta^{AB}\b_B + \frac{r}{2} \delta^{AB}\Big(\frac{a_1}{r_c}\partial_B P +{a_2}\partial^2 v_B -\frac{2}{r_c}f(r)v^D \partial_D \b_B \Big) \nonumber \\
&& + (f^2 g + \frac{3}{2}rfg^2 + \frac{1}{2}r^2 g^3)\delta^{AB}\delta^{CD}\b_B \b_C \b_D - f\delta^{AB}\partial_{\tau}\b_B \nonumber \\
&& -\frac{g}{2} \delta^{AB}\b_{B}\Big[a_3(x^C\Omega_C x^{D}\Omega_D) + a_4(\delta^{CD}\Omega_C\Omega_D \delta_{PQ}x^Px^Q)\Big] \nonumber \\
&& - a_3\delta^{AB}\Omega_B(\Omega_Dx^D) - a_4(\delta^{CD}\Omega_C \Omega_D x^A) +\mathcal{O}(\epsilon^5) \\
\Gamma^{A}_{\tau r} &=& -\frac{1}{2}(f' + g)\delta^{AB}\b_B - \frac{1}{2}\delta^{AB}\Big(\frac{a_1}{r_c}\partial_B P + {a_2}\partial^2 v_B -\frac{2}{r_c}f(r)v^D \partial_D \b_B \Big) \nonumber \\
&& -(fg^2 +\frac{1}{2}r g^3 +\frac{1}{2}f'r g^2 + f'fg)\delta^{AB}\delta^{CD}\b_B \b_C \b_D - \frac{1}{2}g\delta^{AB}\partial_{\tau}\b_B +\mathcal{O}(\epsilon^5) \\
\Gamma^{A}_{\tau B} &=& \frac{1}{2}(ff' + rgf')\delta^{AC}\b_B \b_C + \frac{f}{2}\delta^{AC}(\partial_C \b_B - \partial_B \b_C) +\mathcal{O}(\epsilon^4) \\
\Gamma^{A}_{rr} &=& -g'\delta^{AB}\b_B - (r g^2 g'+ 2fgg')\delta^{AB}\delta^{CD}\b_B \b_C \b_D +\mathcal{O}(\epsilon^5) \\
\Gamma^{A}_{r B} &=& -\frac{1}{2}f'g \delta^{AC}\b_B \b_C + \frac{g}{2}\delta^{AC}(\partial_C \b_B -\partial_B \b_C)  +\mathcal{O}(\epsilon^4)\\
\Gamma^{A}_{BC} &=& -(fg + \frac{1}{2}rg^2)\delta^{AD}\b_D(\partial_B \b_C + \partial_C \b_B) +\mathcal{O}(\epsilon^5)
\end{eqnarray}
\section{The Inverse Metric and the Christoffel symbols of metric choice \rom{2}}\label{appA1}
In the same way we evaluate the inverse metric corresponding to \eqref{metrixproposal1} as a perturbation series over the flat metric $g_{ab}^{(0)}dx^a dx^b$ =  $-r d\tau^2 + 2 d\tau dr + dx_Adx^A$ with the perturbation identified as $H_{ab} dx^a dx^b$ = $-2\b_A f(r) d\tau dx^A -2 \b_A g(r) dr dx^A -\Big(\frac{2a_1}{r_c}\partial_A P_{\text{eff}}+{2 a_2}\partial^2v_A  -\frac{4}{r_c}f(r)v^D \partial_D \b_A\Big)dx^A dr$. The perturbation contains terms of $\mathcal{O}(\epsilon^1)$ and $\mathcal{O}(\epsilon^3)$. In fact all the computation of the inverse metric components as well as the Christoffel connection components for metric choice $\rom{2}$ \eqref{metrixproposal1} can be retrieved from calculations performed on metric choice $\rom{1}$ \eqref{metrixproposal} by setting $a_3 = 0$, $a_4 = 0$ and replacing $P(\tau, x^A)$ by $P_{\text{eff}}(\tau,x^A)$. Hence we list out only the non trivial changes and the remaining components are identical to the ones computed in Appendix \ref{appA}. The changes to be made in the inverse metric components are :
\begin{eqnarray}
g^{\tau A} &=& g \delta^{AB} \b_B + \delta^{AB}\Big(\frac{a_1}{r_c}\partial_B P_{\text{eff}}+ {a_2}\partial^2 v_B -\frac{2}{r_c}f(r)v^D \partial_D \b_B \Big) \nonumber \\ 
&& + (2fg^2 + r g^3)\delta^{AB} \delta^{CD}\b_B \b_C \b_D +\mathcal{O}(\epsilon^5) \\
g^{rr} &=& r + (f^2 + r^2 g^2 +2 rfg)\delta^{AB}\b_A \b_B + \mathcal{O}(\epsilon^4) \\
g^{rA} &=& (f+ rg)\delta^{AB}\b_B + r \delta^{AB}\Big(\frac{a_1}{r_c}\partial_B P_{\text{eff}}+ {a_2}\partial^2 v_B  -\frac{2}{r_c}f(r)v^D \partial_D \b_B \Big) \nonumber \\
&& +(2f^2 g + 3rfg^2 +r^2g^3)\delta^{AB}\delta^{CD} \b_B \b_C \b_D + \mathcal{O}(\epsilon^5) ~.
\end{eqnarray}

Similarly the changes to be made in the Christoffel connection components are,
\begin{eqnarray}
\Gamma^{r}_{\tau \tau} &=& \frac{r}{2} + \frac{1}{2}(f^2 + r^2g^2 + 2rfg)\delta^{AB}\b_A \b_B + \mathcal{O}(\epsilon^4)\\
\Gamma^{r}_{\tau A} &=& \frac{1}{2}r f' \b_A + \frac{1}{2}(f^2f' + r^2 g^2 f' + 2rf'fg)\b_A \delta^{BC}\b_B\b_C \nonumber \\
&& -\frac{r}{2} g \partial_{\tau}\b_A + \frac{1}{2}(f^2 + rfg)\delta^{BC}\b_C(\partial_{B}\b_A -\partial_A \b_B) +\mathcal{O}(\epsilon^5) \\
\Gamma^{A}_{\tau \tau} &=& \frac{1}{2}(f + rg)\delta^{AB}\b_B + \frac{r}{2} \delta^{AB}\Big(\frac{a_1}{r_c}\partial_B P_{\text{eff}}+ a_2\partial^2 v_B  -\frac{2}{r_c}f(r)v^D \partial_D \b_B \Big) \nonumber \\
&& + (f^2 g + \frac{3}{2}rfg^2 + \frac{1}{2}r^2 g^3)\delta^{AB}\delta^{CD}\b_B \b_C \b_D - f\delta^{AB}\partial_{\tau}\b_B  +\mathcal{O}(\epsilon^5)\\
\Gamma^{A}_{\tau r} &=& -\frac{1}{2}(f' + g)\delta^{AB}\b_B - \frac{1}{2}\delta^{AB}\Big(\frac{a_1}{r_c}\partial_B P_{\text{eff}}+ a_2\partial^2 v_B  -\frac{2}{r_c}f(r)v^D \partial_D \b_B \Big) \nonumber \\
&& -(fg^2 +\frac{1}{2}r g^3 +\frac{1}{2}f'r g^2 + f'fg)\delta^{AB}\delta^{CD}\b_B \b_C \b_D - \frac{1}{2}g\delta^{AB}\partial_{\tau}\b_B +\mathcal{O}(\epsilon^5)
\end{eqnarray}
\section{Projected parallel transport equation \eqref{projectedautoparallel} upto $\mathcal{O}(\epsilon^3)$ order for metric choice \rom{1}}\label{appB}
We first consider the projected parallel transport equation with the free index $c = \tau$ \eqref{projautoattau},
\begin{eqnarray}
\gamma_{a \tau}(v^b \nabla_b v^a)|_{r=r_c} &=& \g_{\tau \tau}\Big(\Gamma^{\tau}_{\tau \tau}+ 2\Gamma^{\tau}_{\tau A}v^A + \Gamma^{\tau}_{A B}v^A v^B\Big) + \g_{r \tau}\Big(\Gamma^{r}_{\tau \tau} + 2 \Gamma^{r}_{\tau A} v^A + \Gamma^{r}_{AB}v^A v^B\Big)\nonumber \\ 
&& \g_{B \tau}\Big(\partial_{\tau}v^B + v^A \partial_A v^B + \Gamma^{B}_{\tau \tau} + 2 \Gamma^{B}_{\tau A} v^A + \Gamma^{B}_{A D}v^A v^D\Big)  
\end{eqnarray}
At the following orders of the hydrodynamic expansion parameter, we have the R.H.S as,
\begin{eqnarray}
\mathcal{O}(\epsilon^0) &:\quad  -r_c \Gamma^{\tau(0)}_{\tau \tau} + \Gamma^{r(0)}_{\tau \tau} = 0 ~.
\end{eqnarray}
\begin{eqnarray}
 \mathcal{O}(\epsilon^1) &:\quad \g_{\tau \tau}^{(0)}(\Gamma^{\tau(1)}_{\tau \tau}+2\Gamma^{\tau(0)}_{\tau A}v^A) + \g_{r \tau}^{(0)}(\Gamma^{r(1)}_{\tau \tau} + \Gamma^{r(0)}_{\tau A}v^A) +\g_{B \tau}^{(1)}(\Gamma^{B(0)}_{\tau \tau}) = 0 ~.
\end{eqnarray}
\begin{eqnarray}
\mathcal{O}(\epsilon^2)&:\quad \g_{\tau \tau}^{(0)}(\Gamma^{\tau(2)}_{\tau \tau} + 2 \Gamma^{\tau (1)}_{\tau A}v^A + \Gamma^{\tau (0)}_{AB}v^A v^B) + \g_{r \tau}^{(0)}(\Gamma^{r(2)}_{\tau \tau} + 2 \Gamma^{r(1)}_{\tau A}v^A + \Gamma^{r(0)}_{AB}v^A v^B)\nonumber \\
&+ \g_{B \tau}^{(1)}(\Gamma^{B(1)}_{\tau \tau} + 2 \Gamma^{B(0)}_{A \tau}v^A) + \g_{\tau \tau}^{(2)}(\Gamma^{\tau (0)}_{\tau \tau}) = 0 ~.
\end{eqnarray}
\begin{eqnarray}
\mathcal{O}(\epsilon^3)&:\quad \g_{\tau \tau}^{(0)}(\Gamma^{\tau(3)}_{\tau \tau} + 2 \Gamma^{\tau (2)}_{\tau A}v^A + \Gamma^{\tau (1)}_{AB}v^A v^B) + \g_{r \tau}^{(0)}(\Gamma^{r(3)}_{\tau \tau} + 2 \Gamma^{r(2)}_{\tau A}v^A + \Gamma^{r(1)}_{AB}v^A v^B)\nonumber \\
&+\g_{B \tau}^{(1)}(\Gamma^{B(2)}_{\tau \tau} + 2 \Gamma^{B(1)}_{A \tau}v^A + \Gamma^{B(0)}_{AD}v^A v^D) +\g_{\tau \tau}^{(2)}(\Gamma^{\tau (1)}_{\tau \tau} + \Gamma^{\tau (0)}_{\tau A}v^A) = 0 \,  ~.
\end{eqnarray}

Next, we consider the projected parallel transport equation with the free index $c = r$ \eqref{projauoatr},
\begin{eqnarray}
\gamma_{a r}(v^b \nabla_b v^a)|_{r=r_c} &=& \gamma_{\tau r}\Big(\Gamma^{\tau}_{\tau \tau} + 2\Gamma^{\tau
}_{\tau A}v^A + \Gamma^{\tau}_{A B}v^A v^B\Big) + \gamma_{rr}\Big(\Gamma^{r}_{\tau \tau} + 2\Gamma^{r}_{\tau A}v^A + \Gamma^{r}_{A B}v^A v^B\Big) \nonumber \\
&& \gamma_{A r}\Big(\partial_{\tau} v^A + v^B \partial_B v^A + \Gamma^{A}_{\tau \tau} + 2 \Gamma^{A}_{\tau B}v^B + \Gamma^{A}_{B C}v^B v^C\Big) 
\end{eqnarray}
Expanding them order by order in terms of the hydrodynamic scaling parameter $\epsilon$, we obtain the R.H.S as,
\begin{eqnarray}
\mathcal{O}(\epsilon^0) &:\quad \Gamma^{\tau(0)}_{\tau \tau} -\frac{1}{r_c}(\Gamma^{r(0)}_{\tau \tau}) = 0 ~.
\end{eqnarray}
\begin{eqnarray}
\mathcal{O}(\epsilon^1) &:\quad \g_{\tau r}^{(0)}(\Gamma^{\tau(1)}_{\tau \tau}+2\Gamma^{\tau(0)}_{\tau A}v^A) +\g_{rr}^{(0)}(\Gamma^{r(1)}_{\tau \tau} + 2\Gamma^{r(0)}_{\tau A}v^A)\nonumber\\
& +\g_{B r}^{(1)}(\Gamma^{B(0)}_{\tau \tau}) = 0 ~.
\end{eqnarray}
\begin{eqnarray}
\mathcal{O}(\epsilon^2)&:\quad \g_{\tau r}^{(0)}(\Gamma^{\tau(2)}_{\tau \tau} + 2 \Gamma^{\tau (1)}_{\tau A}v^A + \Gamma^{\tau (0)}_{AB}v^A v^B) +\g_{rr}^{(0)}(\Gamma^{r(2)}_{\tau \tau} + 2 \Gamma^{r(1)}_{\tau A}v^A + \Gamma^{r(0)}_{AB}v^A v^B)\nonumber \\
&+\g_{B r}^{(1)}(\Gamma^{B(1)}_{\tau \tau} + 2 \Gamma^{B(0)}_{A \tau}v^A) + \g_{r r}^{(2)}(\Gamma^{r (0)}_{\tau \tau})\nonumber \\
& = -\frac{1}{2 r_c}\delta^{CD}\b_C \b_D \Big(f^2(r_c) + 2r_c f(r_c)g(r_c) +r_{c}^2 g^2\Big) ~. 
\end{eqnarray}
\begin{eqnarray}
\mathcal{O}(\epsilon^3)&:\quad \g_{\tau r}^{(0)}(\Gamma^{\tau(3)}_{\tau \tau} + 2 \Gamma^{\tau (2)}_{\tau A}v^A + \Gamma^{\tau (1)}_{AB}v^A v^B) +\g_{rr}^{(0)}(\Gamma^{r(3)}_{\tau \tau} + 2 \Gamma^{r(2)}_{\tau A}v^A + \Gamma^{r(1)}_{AB}v^A v^B)\nonumber \\
&+\g_{B r}^{(1)}(\Gamma^{B(2)}_{\tau \tau} + 2 \Gamma^{B(1)}_{A \tau}v^A + \Gamma^{B(0)}_{AD}v^A v^D) + \g_{r r}^{(2)}(\Gamma^{r (1)}_{\tau \tau} + 2 \Gamma ^{r (0)}_{\tau A}v^A) = 0 ~.
\end{eqnarray}
In a similar fashion we consider the projection of the parallel transport equation on the $r = r_c$ cutoff hypersurfave with the free index $c = A$ i.e \eqref{projautoatA}.
\begin{eqnarray}
\gamma_{a A}(v^b \nabla_b v^a)|_{r=r_c} &=&\gamma_{\tau A}\Big(\Gamma^{\tau}_{\tau \tau} + 2 \Gamma^{\tau}_{\tau D} v^D + \Gamma^{\tau}_{CD}v^C v^D \Big) + \gamma_{rA}\Big(\Gamma^{r}_{\tau \tau} + 2 \Gamma^{r}_{\tau D}v^D + \Gamma^{r}_{CD}v^C v^D\Big) \nonumber \\
&& \gamma_{BA}\Big(\partial_{\tau}v^B + v^{C}\partial_{C}v^B + \Gamma^{B}_{\tau \tau}+ 2 \Gamma^{B}_{\tau D}v^D +\Gamma^{B}_{C D}v^C v^D\Big)~.
\end{eqnarray}
Expanding the above equation order by order in terms of $\epsilon$, we have the corresponding set,
\begin{eqnarray}
\mathcal{O}(\epsilon^0)&:\quad \delta_{AB}\Gamma^{B}_{\tau \tau} = 0 ~.
\end{eqnarray}
\begin{eqnarray}
\mathcal{O}(\epsilon^1)&:\quad -f(r_c)\b_A \Gamma^{\tau (0)}_{\tau \tau} - g(r_c)\b_A \Gamma^{r(0)}_{\tau \tau} + \delta_{AB}(\Gamma^{B(1)}_{\tau \tau}+ 2\Gamma^{B(0)}_{\tau D}v^D) = 0 ~.
\end{eqnarray}
\begin{eqnarray}
\mathcal{O}(\epsilon^2)&:\quad \g_{\tau A}^{(1)}(\Gamma^{\tau (1)}_{\tau \tau} + 2 \Gamma^{\tau(0)}_{\tau D}v^D) +\g_{r A}^{(1)}(\Gamma^{r(1)}_{\tau \tau}+ 2\Gamma^{r(0)}_{\tau D}v^D) \nonumber \\
&+\delta_{AB}(\Gamma^{B(2)}_{\tau \tau}+ 2\Gamma^{B(1)}_{\tau D}v^D + \Gamma^{B(0)}_{CD}v^C v^D) = 0 ~.
\end{eqnarray}
\begin{eqnarray}
\mathcal{O}(\epsilon^3)&:\quad \g_{\tau A}^{(1)}(\Gamma^{\tau (2)}_{\tau \tau} + 2 \Gamma^{\tau(1)}_{\tau D}v^D + \Gamma^{\tau(0)}_{CD}v^C v^D)+\g_{r A}^{(1)}(\Gamma^{r(2)}_{\tau \tau}+ 2\Gamma^{r(1)}_{\tau D}v^D + \Gamma^{r(0)}_{CD}v^C v^D) \nonumber \\
&+\delta_{AB}(\partial_{\tau} v^B + v^C\partial_{C}v^B + \Gamma^{B(3)}_{\tau \tau}+ 2\Gamma^{B(2)}_{\tau D}v^D + \Gamma^{B(1)}_{CD}v^C v^D) = \partial_{\tau} v_A + v^C\partial_C v_A + f(r_c)v^C\partial_C\b_A  \nonumber \\
& + \frac{r_c}{2}\Big(\frac{a_1}{r_c}\partial_A P + a_2 \partial^2 v_A - \frac{2}{r_c}f(r_c)v^D\partial_D\beta_A\Big)  + f(r_c)\Big(\partial_A \b_D -\partial_D \b_A \Big)v^D -a_3\Omega_A(\Omega_P x^P)\nonumber \\
&- a_4(\delta^{CD}\Omega_C \Omega_D \delta_{AP}x^{P})
\end{eqnarray}
\section{Projected parallel transport equation \eqref{projectedautoparallel} upto $\mathcal{O}(\epsilon^3)$ order for metric choice \rom{2}}\label{appB1}
We first consider the projected auto-parallel equation with the free index $c = \tau$ \eqref{projautoattau},
\begin{eqnarray}
\gamma_{a \tau}(v^b \nabla_b v^a)|_{r=r_c} &=& \g_{\tau \tau}\Big(\Gamma^{\tau}_{\tau \tau}+ 2\Gamma^{\tau}_{\tau A}v^A + \Gamma^{\tau}_{A B}v^A v^B\Big) + \g_{r \tau}\Big(\Gamma^{r}_{\tau \tau} + 2 \Gamma^{r}_{\tau A} v^A + \Gamma^{r}_{AB}v^A v^B\Big)\nonumber \\ 
&& \g_{B \tau}\Big(\partial_{\tau}v^B + v^A \partial_A v^B + \Gamma^{B}_{\tau \tau} + 2 \Gamma^{B}_{\tau A} v^A + \Gamma^{B}_{A D}v^A v^D\Big)  
\end{eqnarray}
At the following orders of the hydrodynamic expansion parameter, we have the R.H.S as,
\begin{eqnarray}
\mathcal{O}(\epsilon^0) &:\quad  -r_c \Gamma^{\tau(0)}_{\tau \tau} + \Gamma^{r(0)}_{\tau \tau} = 0 ~.
\end{eqnarray}
\begin{eqnarray}
 \mathcal{O}(\epsilon^1) &:\quad -r_c(\Gamma^{\tau(1)}_{\tau \tau}+2\Gamma^{\tau(0)}_{\tau A}v^A) + (\Gamma^{r(1)}_{\tau \tau} + \Gamma^{r(0)}_{\tau A}v^A)\nonumber\\
&-(f(r_c)\b_B)(\Gamma^{B(0)}_{\tau \tau}) = 0 ~.
\end{eqnarray}
\begin{eqnarray}
\mathcal{O}(\epsilon^2)&:\quad -r_c(\Gamma^{\tau(2)}_{\tau \tau} + 2 \Gamma^{\tau (1)}_{\tau A}v^A + \Gamma^{\tau (0)}_{AB}v^A v^B) + (\Gamma^{r(2)}_{\tau \tau} + 2 \Gamma^{r(1)}_{\tau A}v^A + \Gamma^{r(0)}_{AB}v^A v^B)\nonumber \\
&-f(r_c)\b_B(\Gamma^{B(1)}_{\tau \tau} + 2 \Gamma^{B(0)}_{A \tau}v^A) = 0 ~.
\end{eqnarray}
\begin{eqnarray}
\mathcal{O}(\epsilon^3)&:\quad -r_c(\Gamma^{\tau(3)}_{\tau \tau} + 2 \Gamma^{\tau (2)}_{\tau A}v^A + \Gamma^{\tau (1)}_{AB}v^A v^B) + (\Gamma^{r(3)}_{\tau \tau} + 2 \Gamma^{r(2)}_{\tau A}v^A + \Gamma^{r(1)}_{AB}v^A v^B)\nonumber \\
&-f(r_c)\b_B(\Gamma^{B(2)}_{\tau \tau} + 2 \Gamma^{B(1)}_{A \tau}v^A + \Gamma^{B(0)}_{AD}v^A v^D) = 0 \,  ~.
\end{eqnarray}

Next, we consider the projected parallel transport equation with the free index $c = r$ \eqref{projauoatr},
\begin{eqnarray}
\gamma_{a r}(v^b \nabla_b v^a)|_{r=r_c} &=& \gamma_{\tau r}\Big(\Gamma^{\tau}_{\tau \tau} + 2\Gamma^{\tau
}_{\tau A}v^A + \Gamma^{\tau}_{A B}v^A v^B\Big) + \gamma_{rr}\Big(\Gamma^{r}_{\tau \tau} + 2\Gamma^{r}_{\tau A}v^A + \Gamma^{r}_{A B}v^A v^B\Big) \nonumber \\
&& \gamma_{A r}\Big(\partial_{\tau} v^A + v^B \partial_B v^A + \Gamma^{A}_{\tau \tau} + 2 \Gamma^{A}_{\tau B}v^B + \Gamma^{A}_{B C}v^B v^C\Big) 
\end{eqnarray}
Expanding them order by order in terms of the hydrodynamics scaling parameter $\epsilon$, we obtain the R.H.S as,
\begin{eqnarray}
\mathcal{O}(\epsilon^0) &:\quad \Gamma^{\tau(0)}_{\tau \tau} -\frac{1}{r_c}(\Gamma^{r(0)}_{\tau \tau}) = 0 ~.
\end{eqnarray}
\begin{eqnarray}
\mathcal{O}(\epsilon^1) &:\quad -(\Gamma^{\tau(1)}_{\tau \tau}+2\Gamma^{\tau(0)}_{\tau A}v^A) -\frac{1}{r_c}(\Gamma^{r(1)}_{\tau \tau} + 2 \Gamma^{r(0)}_{\tau A}v^A)\nonumber\\
&-(g(r_c)\b_B)(\Gamma^{B(0)}_{\tau \tau}) = 0 ~.
\end{eqnarray}
\begin{eqnarray}
\mathcal{O}(\epsilon^2)&:\quad (\Gamma^{\tau(2)}_{\tau \tau} + 2 \Gamma^{\tau (1)}_{\tau A}v^A + \Gamma^{\tau (0)}_{AB}v^A v^B) -\frac{1}{r_c}(\Gamma^{r(2)}_{\tau \tau} + 2 \Gamma^{r(1)}_{\tau A}v^A + \Gamma^{r(0)}_{AB}v^A v^B)\nonumber \\
&-g(r_c)\b_B(\Gamma^{B(1)}_{\tau \tau} + 2 \Gamma^{B(0)}_{A \tau}v^A) = -\frac{1}{2 r_c}\delta^{CD}\b_C \b_D \Big(f^2(r_c) + 2r_c f(r_c)g(r_c) +r_{c}^2 g^2\Big) ~.  
\end{eqnarray}
\begin{eqnarray}
\mathcal{O}(\epsilon^3)&:\quad (\Gamma^{\tau(3)}_{\tau \tau} + 2 \Gamma^{\tau (2)}_{\tau A}v^A + \Gamma^{\tau (1)}_{AB}v^A v^B) -\frac{1}{r_c} (\Gamma^{r(3)}_{\tau \tau} + 2 \Gamma^{r(2)}_{\tau A}v^A + \Gamma^{r(1)}_{AB}v^A v^B)\nonumber \\
&-g(r_c)\b_B(\Gamma^{B(2)}_{\tau \tau} + 2 \Gamma^{B(1)}_{A \tau}v^A + \Gamma^{B(0)}_{AD}v^A v^D) = 0 ~.
\end{eqnarray}
In a similar fashion we consider the projection of the parallel transport equation on the $r = r_c$ cutoff hypersurfave with the free index $c = A$ i.e \eqref{projautoatA}.
\begin{eqnarray}
\gamma_{a A}(v^b \nabla_b v^a)|_{r=r_c} &=&\gamma_{\tau A}\Big(\Gamma^{\tau}_{\tau \tau} + 2 \Gamma^{\tau}_{\tau D} v^D + \Gamma^{\tau}_{CD}v^C v^D \Big) + \gamma_{rA}\Big(\Gamma^{r}_{\tau \tau} + 2 \Gamma^{r}_{\tau D}v^D + \Gamma^{r}_{CD}v^C v^D\Big) \nonumber \\
&& \gamma_{BA}\Big(\partial_{\tau}v^B + v^{C}\partial_{C}v^B + \Gamma^{B}_{\tau \tau}+ 2 \Gamma^{B}_{\tau D}v^D +\Gamma^{B}_{C D}v^C v^D\Big)~.
\end{eqnarray}
Expanding the above equation order by order in terms of $\epsilon$, we have the corresponding set,
\begin{eqnarray}
\mathcal{O}(\epsilon^0)&:\quad \delta_{AB}\Gamma^{B}_{\tau \tau} = 0 ~.
\end{eqnarray}
\begin{eqnarray}
\mathcal{O}(\epsilon^1)&:\quad -f(r_c)\b_A \Gamma^{\tau (0)}_{\tau \tau} - g(r_c)\b_A \Gamma^{r(0)}_{\tau \tau} + \delta_{AB}(\Gamma^{B(1)}_{\tau \tau}+ 2\Gamma^{B(0)}_{\tau D}v^D) = 0 ~.
\end{eqnarray}
\begin{eqnarray}
\mathcal{O}(\epsilon^2)&:\quad -f(r_c)\b_A(\Gamma^{\tau (1)}_{\tau \tau} + 2 \Gamma^{\tau(0)}_{\tau D}v^D) - g(r_c)\b_A (\Gamma^{r(1)}_{\tau \tau}+ 2\Gamma^{r(0)}_{\tau D}v^D) \nonumber \\
&+\delta_{AB}(\Gamma^{B(2)}_{\tau \tau}+ 2\Gamma^{B(1)}_{\tau D}v^D + \Gamma^{B(0)}_{CD}v^C v^D) = 0 ~.
\end{eqnarray}
\begin{eqnarray}
\mathcal{O}(\epsilon^3)&:\quad -f(r_c)\b_A(\Gamma^{\tau (2)}_{\tau \tau} + 2 \Gamma^{\tau(1)}_{\tau D}v^D + \Gamma^{\tau(0)}_{CD}v^C v^D)\nonumber \\
 &- g(r_c)\b_A (\Gamma^{r(2)}_{\tau \tau}+ 2\Gamma^{r(1)}_{\tau D}v^D + \Gamma^{r(0)}_{CD}v^C v^D) \nonumber \\
&+\delta_{AB}(\partial_{\tau} v^B + v^C\partial_{C}v^B + \Gamma^{B(3)}_{\tau \tau}+ 2\Gamma^{B(2)}_{\tau D}v^D + \Gamma^{B(1)}_{CD}v^C v^D) = \partial_{\tau} v_A + v^C\partial_C v_A \nonumber \\
 &+ \frac{r_c}{2}\Big(\frac{a_1}{r_c}\partial_A P_{\text{eff}}+ {a_2}\partial^2v_A  - \frac{2}{r_c}f(r_c)v^D\partial_D\beta_A\Big)\nonumber \\
 &- f(r_c) \partial_{\tau}\b_A + f(r_c)\Big(\partial_A \b_D -\partial_D \b_A \Big)v^D ~.
\end{eqnarray}

\section{Calculation of expansion scalar $\Theta$ \eqref{theta2} upto $\mathcal{O}(\epsilon^3)$ order for choice \rom{1}}\label{appC} 
The expansion scalar as derived on the cut-off hypersurface $r =r_c$ has the following form,
\begin{eqnarray}
\Theta &=& \underbrace{-\g^{\tau \tau} \Big(\Gamma^{\tau}_{\tau \tau}v_{\tau} + \Gamma^{r}_{\tau \tau}v_{r} + \Gamma^{A}_{\tau \tau}{v}_{A} \Big)}_\text{expression 1}+\underbrace{-2 \g^{\tau A}\Big(\Gamma^{\tau}_{\tau A}v_{\tau}+\Gamma^{r}_{\tau A}v_r + \Gamma^{A}_{\tau A}{v}_A\Big)\nonumber}_\text{expression 2} \\
&& + \underbrace{\g^{AB}\Big(\partial_A {v}_B -\Gamma^{\tau}_{AB}v_{\tau} - \Gamma^{r}_{AB}v_r + \Gamma^{D}_{AB}{v}_D\Big)}_\text{ expression 3}
\label{theta3}
\end{eqnarray}
Evaluating "expression $1$" from \eqref{theta3}, with the relevant $\Gamma$'s obtained in Appendix \ref{appA} we find its leading order to be $\mathcal{O}(\epsilon^4)$. Similarly, the leading order contribution in "expression $2$" hits at $\mathcal{O}(\epsilon^4)$. The leading order behaviour of "expression $3$" occurs at $\mathcal{O}(\epsilon^2)$ and is equivalent to $\partial_A v^A$ and it vanishes at $\mathcal{O}(\epsilon^3)$.
\section{Calculation of expansion scalar $\Theta$ \eqref{theta21} upto $\mathcal{O}(\epsilon^3)$ order for choice \rom{2}}\label{appC1} 
The "expansion parameter" as derived on the cut-off hypersurface $r =r_c$ had the following form,
\begin{eqnarray}
\Theta &=& \underbrace{\frac{1}{r_c} \Big(\Gamma^{\tau}_{\tau \tau}v_{\tau} + \Gamma^{r}_{\tau \tau}v_{r} + \Gamma^{A}_{\tau \tau} \tilde{v}_{A} \Big)}_\text{expression 1}+\underbrace{\Big(2 \frac{f(r_c)}{r_c}\delta^{AB}\b_B\Big)\Big(\Gamma^{\tau}_{\tau A}v_{\tau}+\Gamma^{r}_{\tau A}v_r + \Gamma^{A}_{\tau A}\tilde{v}_A\Big)\nonumber}_\text{expression 2} \\
&& + \underbrace{\delta^{AB}\Big(\partial_A \tilde{v}_B -\Gamma^{\tau}_{AB}v_{\tau} - \Gamma^{r}_{AB}v_r + \Gamma^{D}_{AB}\tilde{v}_D\Big)}_\text{ expression 3}
\label{theta31}
\end{eqnarray}
Evaluating "expression $1$" from \eqref{theta31}, with the relevant $\Gamma$'s obtained in Appendix \ref{appA} we find its leading order to be $\mathcal{O}(\epsilon^4)$. Similarly, the leading order contribution in "expression $2$" hits at $\mathcal{O}(\epsilon^4)$. The leading order behaviour of "expression $3$" occurs at $\mathcal{O}(\epsilon^2)$ and is equivalent to $\partial_A v^A$ and it vanishes at $\mathcal{O}(\epsilon^3)$.


\end{document}